\documentclass[aps,prl,twocolumn,groupedaddress, nobibnotes]{revtex4-2}

\usepackage{amsthm}
\usepackage{mathtools}
\usepackage{physics}
\usepackage{graphicx}
\usepackage[left=23mm,right=13mm,top=35mm,columnsep=15pt]{geometry} 
\usepackage{adjustbox}
\usepackage{placeins}
\usepackage[T1]{fontenc}
\usepackage{lipsum}
\usepackage{csquotes}

\usepackage[english]{babel}
\usepackage[utf8]{inputenc}
\usepackage[colorinlistoftodos, color=green!40, prependcaption]{todonotes}

\usepackage[pdftex, pdftitle={Article}, pdfauthor={Author}]{hyperref} 
\usepackage{hypcap}
\usepackage{bm}
\usepackage{cleveref}
\usepackage[markup=default, authormarkup=none]{changes}
\definechangesauthor[name={Yifan Liu}, color=black]{YF}

\begin{document}


\title{Tunable Coupling, Topology, and Chirality by Antimagnons in Magnetic Multilayer}

\author{Yifan Liu$^{1}$, Zehan Chen$^{1,2}$, Qiming Shao$^{1,2}$}
\affiliation{\centering%
$^1$Department of Electronic and Computer Engineering, \mbox{The Hong Kong University of Science and Technology, Hong Kong SAR}\\
$^2$Department of Physics, \mbox{The Hong Kong University of Science and Technology, Hong Kong SAR}}

\date{\today}

\begin{abstract}
Realizing novel topological states in magnonic systems \added[id=YF]{unlocks} robust, low-power spin-wave devices.
In this letter, we show that incorporating left-handed spin waves (antimagnons) fundamentally reorganizes band topology, and enables tunable spin-wave coupling and chirality.
We proposed a two-dimensional Su–Schrieffer–Heeger like model, the 2D-SSH4 chain, where dipolar interactions between magnons and antimagnons generate topological bands with nonzero Chern numbers.  
This framework explains the origin of topological surface states in ferromagnetic multilayer and shows they share the same topological origin as classic magnetostatic surface spin waves. 
Our model also offers a straightforward framework for designing more complex magnetic multilayer connected by dipolar interactions, such as antiferromagnetic/ferromagnetic multilayer. 
\added[id=YF]{In these dipolar-coupled multilayers, both coherent and dissipative interlayer spin-wave couplings together with the layer-resolved chirality, are tunable via external magnetic fields and spin torques.}
Our results provide a practical platform for topological magnonics, enabling control of magnon chirality and coupling in future devices.
\end{abstract}


\maketitle

\let\oldaddcontentsline\addcontentsline  
\renewcommand{\addcontentsline}[3]{}     

\label{sec:Introduction}
    Magnonics has emerged as a promising platform for developing low-power computing \cite{Magnon, low_power_computing} and memory devices \cite{memory_device_1_Shao}, where spin waves, the collective excitations of magnetic order, carry the information. These excitations exhibit two distinct chiralities, namely right-handed (RH) and left-handed (LH) spin waves, whose associated quasiparticles are known as magnons and antimagnons, respectively. Recent research in fermionic topological insulators\cite{fermionic_topo_1, fermionic_topo_2, fermionic_topo_3} inspires the study of magnonic topological insulators (MTIs). These systems host topologically protected edge or surface states that are robust against phonon-induced and impurity-induced backscattering\cite{skyrmion_1, skyrmion_2, skyrmions_phonon, mag_topo_2_zero_freq, mag_topo_3_first_theory, mag_topo_4, mag_topo_5, mag_topo_6_review, Basic_Review, mag_topo_8}.
    \added[id=YF]{While extensive studies devoted to MTIs, topological states involving antimagnons remain largely unexplored because in magnetic materials, the Gilbert damping causes LH spin waves to decay into RH ones. 
    However, magnon states can interact with antimagnons, through an enlarged Hamiltonian that includes both sectors and their cross-terms, unlocking spin-wave states unattainable in magnon-only models \cite{Antimagnonics}. Additionally, manipulation of coherent and dissipative coupling brought by antimagnon and mediated by dipolar interaction allows researchers to engineer nonreciprocal devices such as isolators and circulators \cite{dissipative_1, dissipative_2}.
    Furthermore, the distinct degrees of freedom provided by LH and RH spin waves\cite{Chirality2_comments, Antimagnonics} can benefit applications in information processing\cite{Chiral_device_review, shao2021roadmap} and hybrid systems\cite{hybrid_magnon_review, Hybrid_magnonics, skyrmions_phonon, magnon_magnon_1,photon_magnon_1, photon_magnon_2,magnon_magnon_2_longrange_entangle}, significantly broadening the scope of magnonic applications.}

    In this letter, \added[id=YF]{we investigate the effects of long-range dipolar interaction on the FM multilayer using second quantization\cite{SSH1_Luqiao, Dipolar_1, Dipolar_2, dipolar_multilayer_1_FM, dipolar_multilayer_2_AFM, dipolar_multilayer_3_AFM}. }
    These interactions break both chiral (sublattice) symmetry (CS) and time-reversal symmetry (TRS), giving rise to nontrivial Su–Schrieffer–Heeger (SSH)-like topological states \cite{SSH1_testbook, SSH3_origin}. We name this model the 2D-SSH4 chain \cite{SSH4Model_1, SSH4Model_2, SSH2_India}, in which the magnonic band structures exhibit nonzero Chern numbers.
    We discover that antimagnon states act as intermediate states, hybridizing otherwise disconnected magnon states in different layers. Dissipative magnon-antimagnon coupling via dipolar interactions can transform an originally trivial state into a topologically nontrivial one, which shares the same topological origin as the classic magnetostatic surface spin waves (MSSWs) in FM insulators interestingly.
    \added[id=YF]{Based on this model, we further demonstrate that the interlayer coherent and dissipative spin-wave coupling together with the resulting layer-resolved chirality can be tuned by external magnetic fields and spin–orbit torques (SOTs) \cite{SOT1, SOT2, SOT3_formalism, SOT4_formalism}.}
    \added[id=YF]{Finally, our framework can be straightforwardly generalized to more complex magnetic multilayer, such as antiferromagnetic/ferromagnetic (AFM/FM) multilayer. 
    Synthesizing the features of both parallel and antiparallel FM multilayer yields two coupled 2D-SSH4 chains that well capture the topological surface state of AFM/FM multilayer and reveal rich possibilities for tunable coupling and chirality.}

\label{sec:develop}
    
    \begin{figure}[t] 
    \capstart
    \centering 
    \includegraphics[width=\columnwidth]{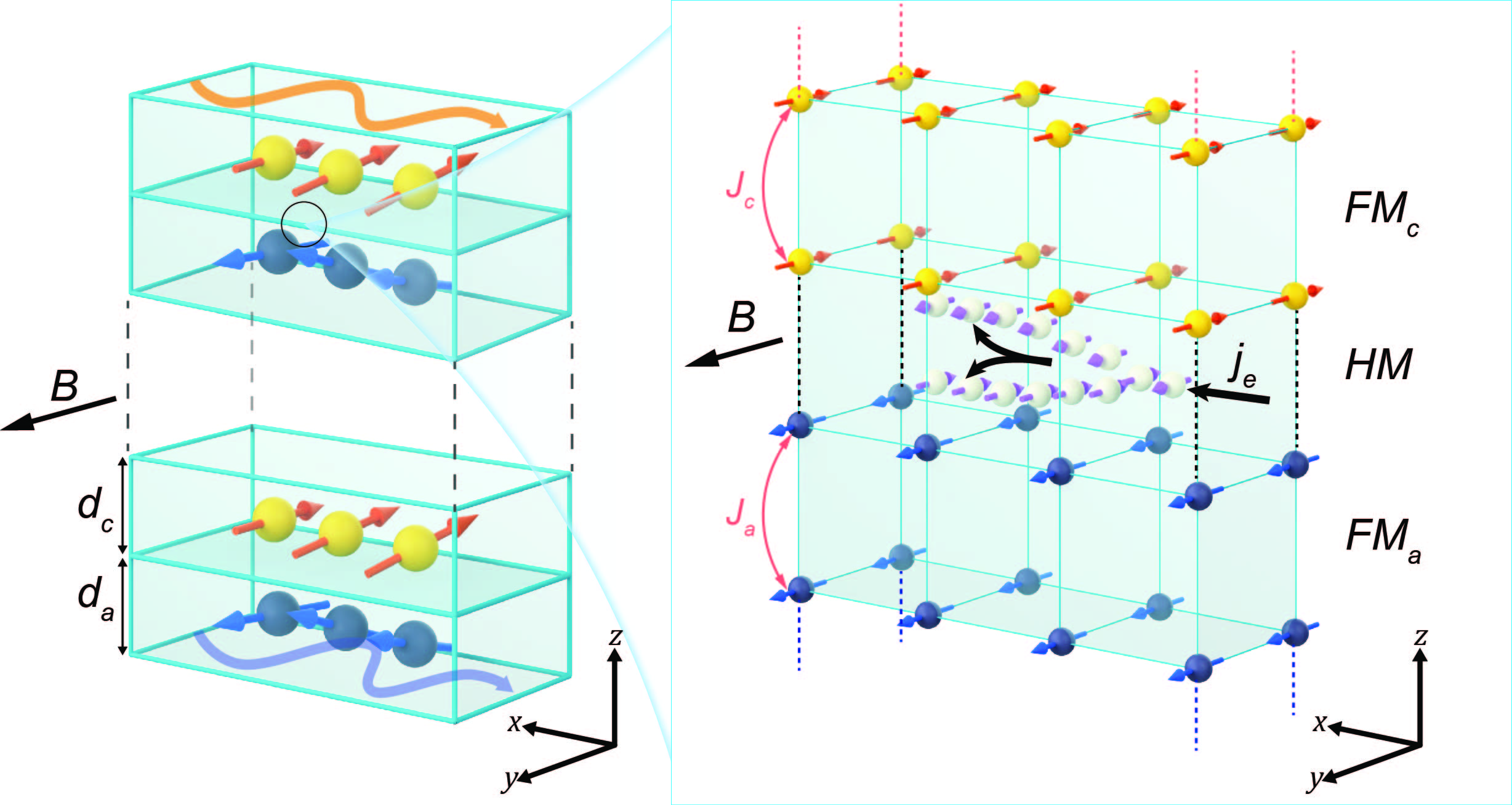} 
    \caption{\label{fig:1}Model of antiparallel FM multilayer. The navy blue and orange arrows are spin moments of atom \(a\) and \(c\), pointing in \(\pm y\) direction. There is a heavy metal (HM) layer between each FM layer with electric current $j_e$ to provide \added[id=YF]{SOT} indicated by purple spin arrow. Each FM layer with the thickness of $d_{a(c)}$ are simple cubic lattice (lattice constant $a_0$) with Heisenberg exchange constant $J_{a(c)}$ between nearest neighbors.}
    \end{figure}

    \textit{2D-SSH4 chain and FM multilayer.---} 
    We revisit the FM multilayer with dipole-exchange spin wave shown in Fig. \ref{fig:1}, which was first proposed by Hu et.al \cite{SSH1_Luqiao}. It consists of \(M\) FM bilayers. Each bilayer is a \added[id=YF]{$z$-direction unit cell}, in which the two different layers are denoted as \(a\) and \(c\). \added[id=YF]{In the following discussion, we consider layer \(a\) as yttrium iron garnet (YIG) and layer \(c\) as permalloy (Py) \cite{YIG/Py_1}.} We can write the Hamiltonian of the bilayer unit cell as:
    \begin{equation}
    \label{eq:H_total}
    \hat{\mathcal H}_{\mathrm{FF}}
       = \sum_{l=a,c}
           \Bigl(
             \hat{\mathcal H}_{l\mathrm{0}}
             + \hat{\mathcal H}_{l\mathrm{D0}}
           \Bigr)
         + \hat{\mathcal H}_{ac\mathrm{D}},
    \end{equation}  
    with
\begin{align}
\label{eq:FFHl0}
\hat{\mathcal H}_{l\mathrm{0}} &=
  -g\mu_{\mathrm B}\sum_{i}\,
     \hat{\bm{S}}_{l i}\!\cdot\!\bm{B}_{l}
  +J_{l}\!\!\sum_{\langle i,j\rangle}
     \hat{\bm{S}}_{l i}\!\cdot\!\hat{\bm{S}}_{l j},
\end{align}
\begin{align}
\hat{\mathcal H}_{l\mathrm D0} &=
  \frac{\mu_{0}\bigl(g\mu_{\mathrm B}\bigr)^{2}}{2}
  \sum_{i\neq j}
    \frac{
      R_{ij}^{2}\,\hat{\bm{S}}_{l i}\!\cdot\!\hat{\bm{S}}_{l j}
      -3\bigl(\bm{R}_{ij}\!\cdot\!\hat{\bm{S}}_{l i}\bigr)
           \bigl(\bm{R}_{ij}\!\cdot\!\hat{\bm{S}}_{l j}\bigr)}
         {R_{ij}^{5}},
\end{align}
\begin{align}
\hspace{-0.7cm}
\hat{\mathcal H}_{ac\mathrm D} &=
  \frac{\mu_{0}\bigl(g\mu_{\mathrm B}\bigr)^{2}}{2}
  \sum_{i\in a, }\sum_{j\in c}
    \frac{
      R_{ij}^{2}\,\hat{\bm{S}}_{a i}\!\cdot\!\hat{\bm{S}}_{c j}
      -3\bigl(\bm{R}_{ij}\!\cdot\!\hat{\bm{S}}_{a i}\bigr)
           \bigl(\bm{R}_{ij}\!\cdot\!\hat{\bm{S}}_{c j}\bigr)}
         {R_{ij}^{5}},
\end{align}
    where  
    \(\hat{\mathcal H}_{l\mathrm{0}}\) contains the intralayer Heisenberg
    exchange and Zeeman interaction,  
    \(\hat{\mathcal H}_{l\mathrm{D0}}\) accounts for the intralayer dipolar
    interaction, and  
    \(\hat{\mathcal H}_{ac\mathrm{D}}\) describes the interlayer dipolar coupling
    between layers \(a\) and \(c\).

    By orienting the spin moments in each layer in the $\pm y$ direction and assuming that they are uniform along $z$ direction, we can express the states in \(M\) unit cells as a tensor product of the magnonic Fock state in each layer and the state of the layer. Adopting periodic boundary conditions in the $z$ direction, we arrive at a 4 by 4 Hamiltonian of the total system. $\mathbf{H}_{\mathrm{FF}}$ and $\mathbf{H}'_{\mathrm{FF}}$ refer to the case when magnetization (the static component of spin moments) in neighboring layers is antiparallel or parallel along the $y$ direction. The detailed expressions are shown in Sec. A of the Supplemental Material \cite{SM}. 

    \begin{figure}[b] 
    \capstart
    \centering 
    \includegraphics[width=\linewidth]{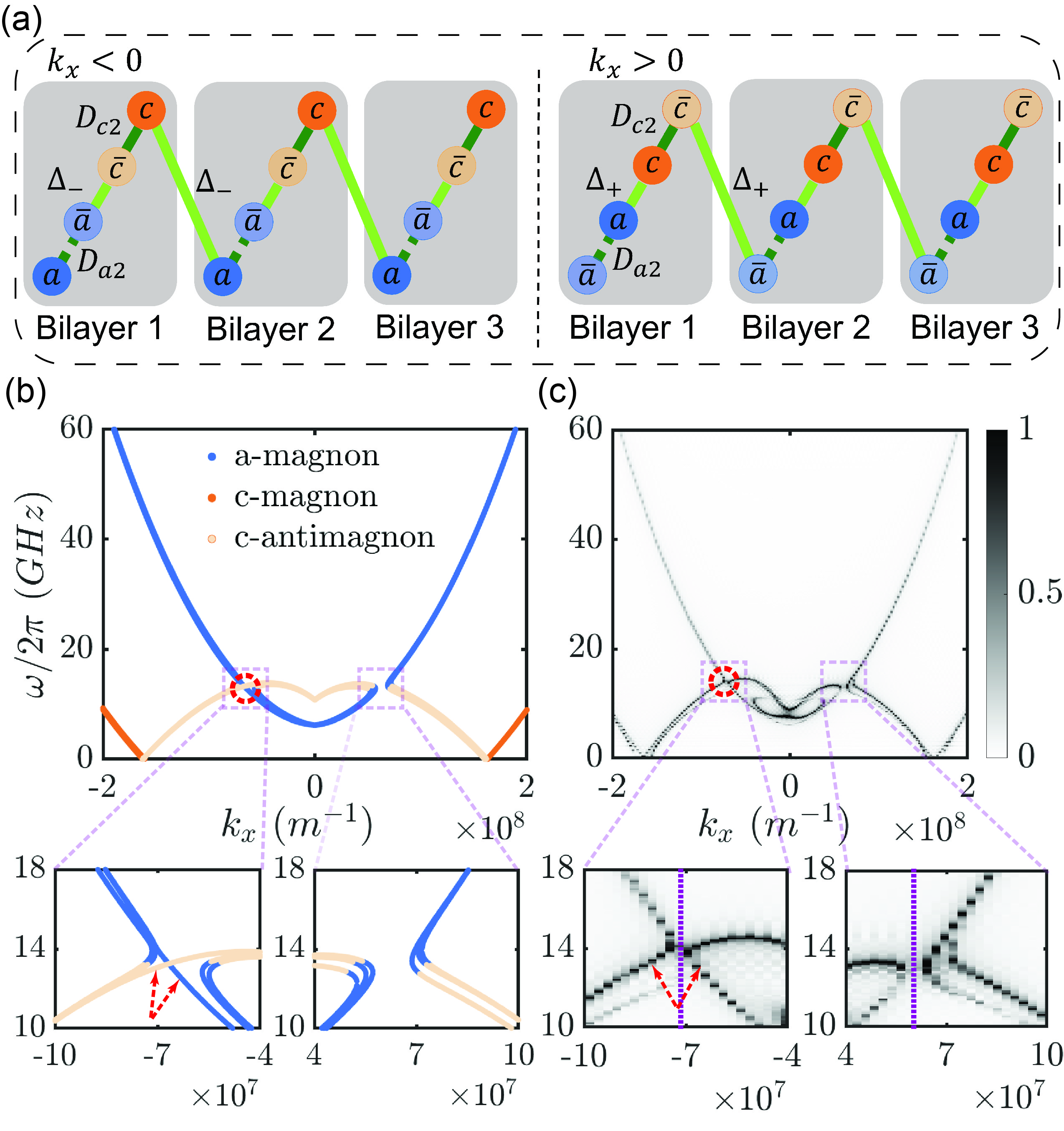} 
    \caption{\label{fig:2}(a)Illustration of the 2D-SSH4 chian for antiparallel case. Magnon (denoted as $a, c$) and antimagnon (denoted as $\bar{a}, \bar{c}$) states of the layers are coupled via interlayer (light green, $\Delta_{\pm}$) and intralayer (dark green, $D_2$) dipolar interactions, with the much weaker intralayer interaction ($D_{a2}$) indicated by dash dark green lines. (b) Calculated band structure for three antiparallel FM-bilayer unit cells; (c) Simulated counterpart for two unit cells, both obtained under the overall effective external field: $B_a^y/\mu_0 = 1.2\times 10^5  \,\mathrm{A/m}$ and $B_c^y/\mu_0 = 8.5\times 10^5\,\mathrm{A/m}$ (nontrivial). Red circles and red arrows indicate nonreciprocal topological surface states \added[id=YF]{between the level-attraction anticrossing bulk states}, and the purple dashed line marks additional simulation points corresponding to Fig. \ref{fig:7}(b) in Sec. D of the Supplemental Material \cite{SM}.}
    \end{figure}

\begin{figure}[t] 
    \capstart
    \centering 
    \includegraphics[width=\linewidth]{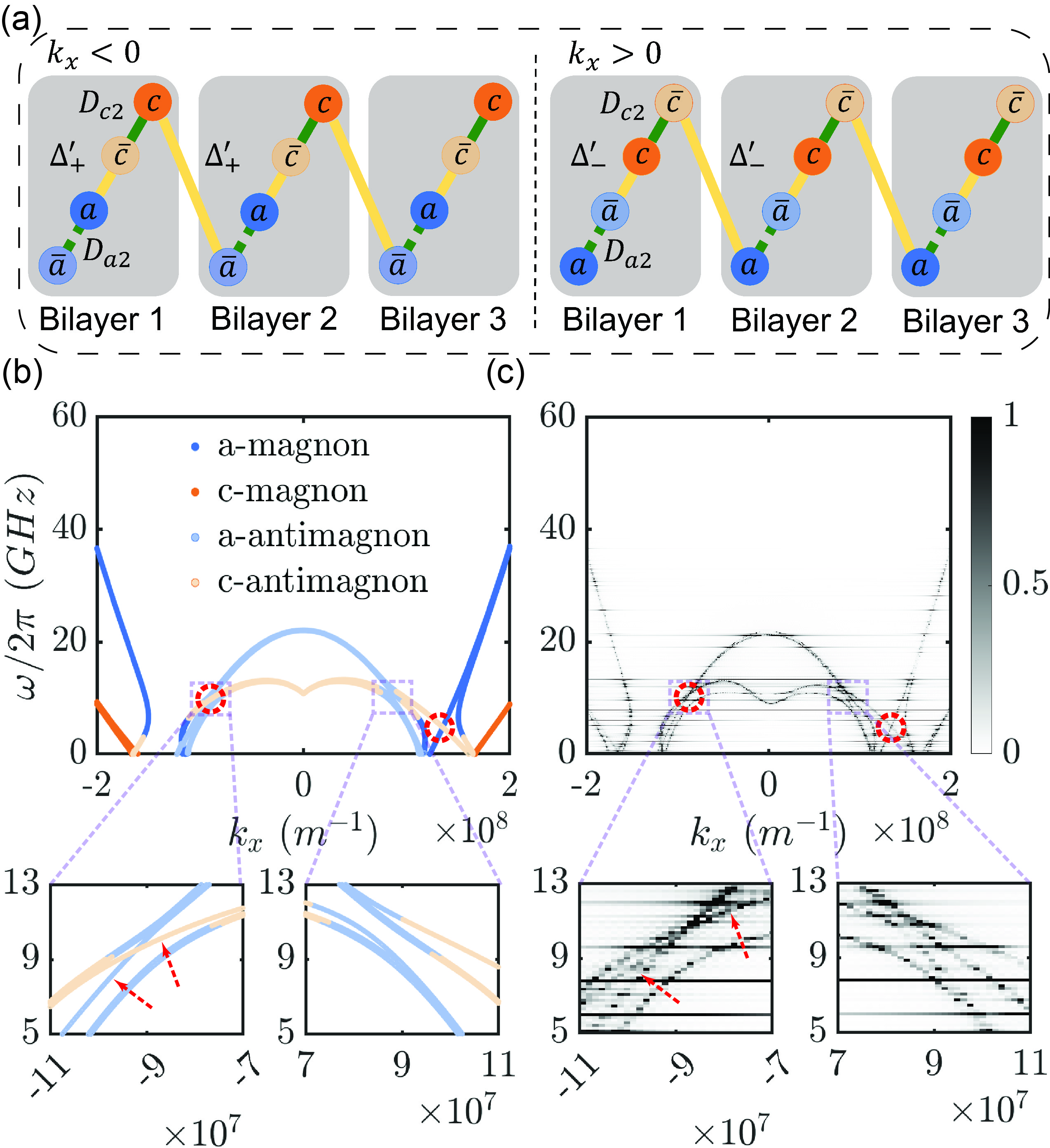} 
    \caption{\label{fig:3}(a)Illustration of the 2D-SSH4 chian for parallel case. Magnon ($a, c$) and antimagnon ($\bar{a}, \bar{c}$) states of the layers are coupled via interlayer (gold, $\Delta'_{\pm}$) and intralayer (dark green, $D_2$) dipolar interactions, with the much weaker intralayer interaction ($D_{a2}$) indicated by dash dark green lines. (b) Calculated band structure for three antiparallel FM-bilayer unit cells; (c) Simulated counterpart for two unit cells, both obtained under the overall effective external field: $B_a^y/\mu_0 = -7\times 10^5  \,\mathrm{A/m}$ and $B_c^y/\mu_0 = -8.5\times 10^5  \,\mathrm{A/m}$ (nontrivial). Red circles and arrows indicate nonreciprocal topological surface states, \added[id=YF]{with the zoomed-in plot shows one of them between the level-repulsive anticrossing bulk states}}
    \end{figure}

    By considering dipolar interactions of traveling spin waves up to the nearest neighbor (which decay exponentially across layers\cite{SSH1_Luqiao}), we obtain the 2D-SSH4 chain shown in Figs.\ref{fig:2}(a) and \ref{fig:3}(a). In the antiparallel case, interlayer dipolar interactions occur only between spin waves of the same chirality (magnon-magnon or antimagnon-antimagnon), while in the parallel case, they arise only between opposite chirality (magnon-antimagnon). \added[id=YF]{In addition, the nonreciprocal dipolar interaction connects different interlayer states for positive and negative $k_x$.} Combined with intralayer dipolar interactions, the multilayer forms an SSH4 chain along the $z$ direction. Considering spin waves only propagating along the $x$ direction in each layer, both the intrinsic frequency and dipolar interaction strengths depend on the wave vector $k_x$. Consequently, the FM multilayer realizes a 2D-SSH4 chain, characterized by the Chern number.

   The topological characterization of the 2D-SSH4 chain is complex. Here we focus on specific situations. In the antiparallel case, the calculated and simulated band structures are shown in Fig. \ref{fig:2}(b)-(c). The darker color represents the magnon-dominating states, while the lighter color represents the antimagnon-dominating ones. Because the intralayer dipolar interaction in layer \(a\) (represented by a dashed dark green bond in Figs. \ref{fig:2}(a)) is much weaker than in layer \(c\) (represented by a solid dark green bond). The nonreciprocal topological surface states, highlighted with red circles in Fig. \ref{fig:2}(b)-(c), emerge when intrinsic band of the head of chain (Bilayer 1 \(a\) for $k_x<0$) crosses the one of tail of chain (Bilayer 3 $\bar{c}$ and \(c\) for $k_x<0$). They are localized at the boundary of the multilayer. Detailed band structure calculations are in Sec. A of the Supplemental Material \cite{SM}. \added[id=YF]{The coupling mechanisms governing the bulk bands are twofold: the magnon-magnon coupling is coherent coupling, characterized by the level-repulsive anticrossing of the bulk band. Meanwhile, the magnon-antimagnon coupling is dissipative coupling, characterized by the level-attraction anticrossing of bulk band \cite{dissipative_1, dissipative_2, dissipative_3, Hybrid_magnonics}. In Sec. A of the Supplemental Material \cite{SM}, the physical origin of the coherent and dissipative coupling is discussed.}

    While $\mathbf{H}_{\mathrm{FF}}$ preserves particle–hole symmetry \cite{mag_topo_2_zero_freq,Basic_Review}, 
    the dipolar interaction connects the magnon and antimagnon subspace, breaking CS
    \cite{Break_Chiral_1,Break_Chiral_2} and TRS \cite{Break_TRS_1,Break_TRS_2}. 
    Therefore, we can use the Chern number to characterize it. 
    However, since our system is bosonic, the Chern number is calculated as
$    \mathrm{Ch}_m=\frac{1}{2\pi}
    \int_{\mathrm{BZ}}\!\mathrm{d}k_x\,\mathrm{d}k_z\,
    \Omega_{m}^{y}(k_x,k_z),$
    where
$    \Omega_{m}^{y}(k_x,k_z)=
     i\partial_{k_x}\langle\chi_{m}|\bm{\eta}\,\partial_{k_z}|\chi_{m}\rangle
     -i\partial_{k_z}\langle\chi_{m}|\bm{\eta}\,\partial_{k_x}|\chi_{m}\rangle,$
    and $\bm{\eta}=\mathrm{diag}(1,1,-1,-1)$ and $|\chi_{m}\rangle$ is the $m^{\text{th}}$ eigenvector of $\mathbf{H}_{\mathrm{FF}}$. 
    The detailed derivation and numerical methods are shown in Sec. A of the Supplemental Material \cite{SM}. 
    The nonzero Chern number will be $1$ or $-1$, indicating a pair of chiral surface states.

    \added[id=YF]{Antimagnon state represents the LH spin wave, indicating a non-equilibrium state under an external magnetic field due to Gilbert damping classically. To sustain such an unstable state \cite{Antimagnonics},} we need an additional \added[id=YF]{SOT} which can be generated by an electrical current in a thin heavy metal (HM) layer between each FM layer \cite{SOT3_formalism, SOT4_formalism}, shown in Fig. \ref{fig:1}. HM also suppresses the AFM exchange coupling between neighboring layers.\cite{Dipolar_1, Dipolar_2}. By tuning the external magnetic field $B$ and electric current $j_e$, we have the freedom to tune the coherent and dissipative coupling by selecting which band to cross and transit between the trivial and nontrivial states. This setup allows the frequency to be tuned to the $\mathrm{GHz}$ range, making it accessible for experimental detection using microwave systems\cite{mag_topo_2_zero_freq, mag_topo_6_review, Basic_Review, challenge_microwave_1, challenge_microwave_2}. To further validate our models, in Sec. D of the Supplemental Material \cite{SM}, we conduct micromagnetic simulations of propagating spin-wave spectroscopy to analyze the system's response to dynamic magnetic field excitation and demonstrate its feasibility of detecting such a state in this experiment.

    In the parallel case, the 2D-SSH4 chain of the FM multilayer is presented in Figs. \ref{fig:3}(a). Similar to the antiparallel configuration, the system supports nonreciprocal topological surface states, which appear when the intrinsic band of the head of the chain (Bilayer 1 \(a\) for $k_x>0$) crosses the intrinsic band of the tail of the chain (Bilayer 3 $\bar{c}$ and \(c\) for $k_x>0$). These surface states are marked by red circles in Fig. \ref{fig:3}(b)-(c), and verified by micromagnetic simulation\cite{COMSOL1_MM, COMSOL2_acdc}. 
    The antimagnon states connect different layers' magnon states by dipolar interaction even without being lifted to positive frequency. Therefore, surface states show up even when two magnon band structures have crossing points in Fig. \ref{fig:3}(c). This phenomenon cannot be explained if we use the conventional magnon-only framework. It provides the accessibility to achieve topological surface state in a more easily constructed FM multilayer than the antiparallel configuration. Also, this non-reciprocity reminds us of the classic MSSW. In the Sec. B of the Supplemental Material \cite{SM}, we show that MWWS has the same topological origin as the 2D-SSH model when breaking the TRS by an external magnetic field.

\label{sec:Tunable Coupling and Chirality}
    
    \begin{figure}[t] 
    \capstart
    \centering 
    \includegraphics[width=\columnwidth]{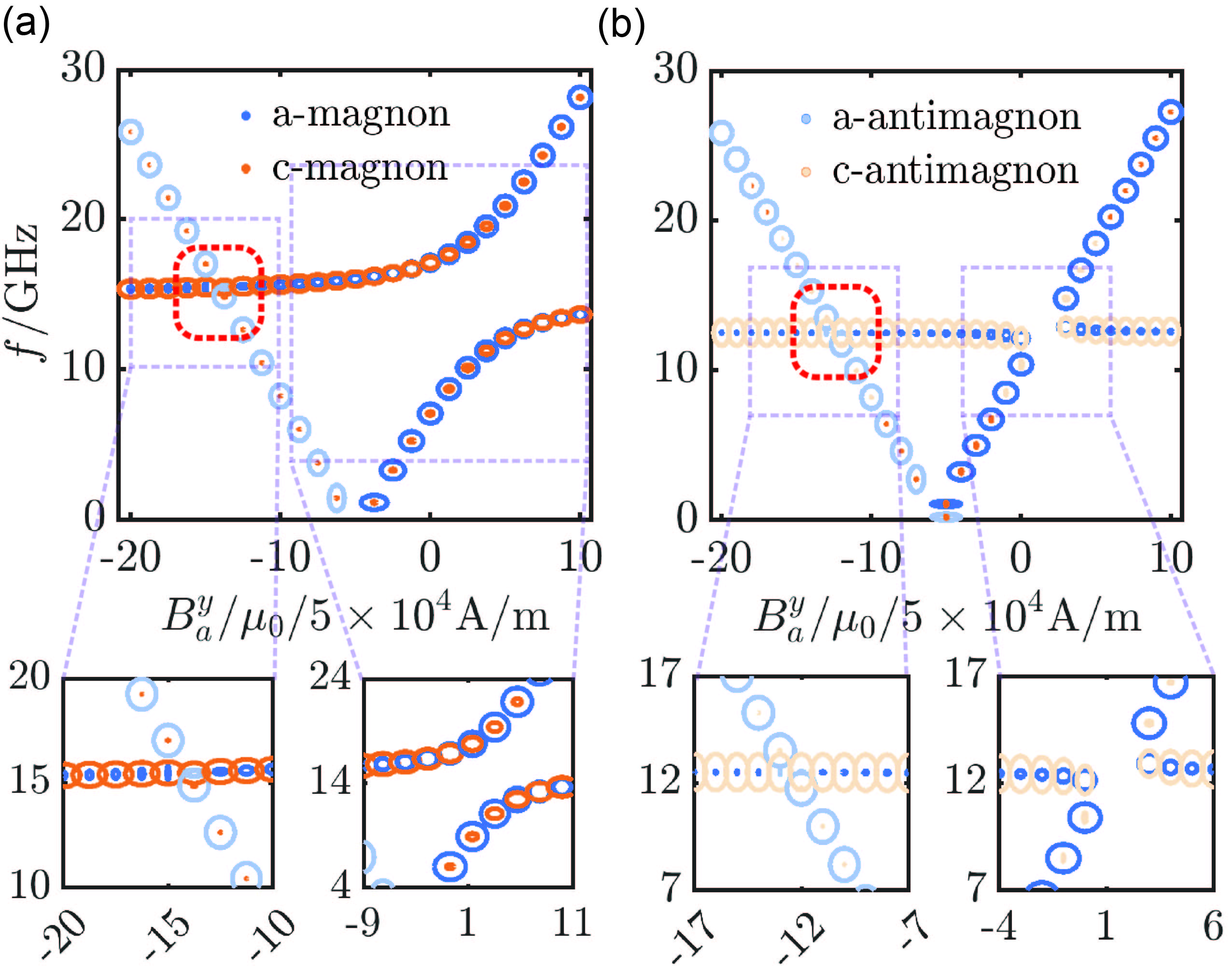} 
    \caption{\label{fig:4.1 coupling} Tunable coupling and chirality in an antiparallel FM bilayer for traveling spin wave at $k_x=6.8\times10^7 \ \mathrm{1/m}$. (a) $B_c^y/\mu_0 = 0\ \mathrm{A/m}$ (b)$B_c^y/\mu_0 = -8.5\times 10^5 \mathrm{A/m}$. Spin wave chirality is shown as ellipses. There are two types of interlayer dipolar coupling, where spin waves hybridize: (a) level-repulsive anticrossings as coherent (same chirality) coupling, whereas (b) level-attractive anticrossings as dissipative (opposite chirality) coupling. Red dashed boxes highlight intrinsic surface states of each uncoupled layer – degenerate modes with no observable splitting.}
    \end{figure}

    \textit{Tunable coupling and chirality.---}  
     \added[id=YF]{Next, we demonstrate the tunability of interlayer spin-wave coupling and chirality in an antiparallel FM bilayer unit.} In Fig. \ref{fig:4.1 coupling}, we plot the trajectories of the magnetization dynamics in layer $a$ and $c$ as an indication of spin-wave chirality across different overall effective external fields of layer $B_a^y/\mu_0$ and fixed $B_c^y/\mu_0$. \added[id=YF]{The ellipticity of each trajectory conveys its net chirality as a superposition of RH and LH spin waves, with darker color meaning magnons dominating and lighter color meaning antimagnons dominating. We depict blue for layer $a$ and orange for layer $c$, respectively. The size represents and spin wave magnitude.} The linearly polarized spin wave emerges near zero frequency, where dissipative coupling occurs between the magnon and antimagnon within the same layer. The coherent (Fig. \ref{fig:4.1 coupling}a) and dissipative (Fig. \ref{fig:4.1 coupling}b) coupling can be tuned by adjusting the external magnetic field to bring spin waves of the same or opposite chirality into frequency resonance. However, since the spin waves propagate along the $+x$ direction, the antimagnon state in layer $a$ becomes almost decoupled from the other modes, as illustrated in Fig. \ref{fig:2}a. This leads to the emergence of intrinsic surface states, highlighted by the red dashed box. In this regime, degenerate spin wave modes from the two layers coexist without hybridization. We utilize the coupled and decoupled regimes to further control the chirality of the spin wave in the bilayer. In particular, we achieve all four possible combinations of chirality in the FM bilayer unit under the same external fields but with different $k_x$, with the mixed-handed state protected by the decoupled surface modes, discussed in Sec. D of the Supplemental Material \cite{SM}.

\begin{figure}[t] 
    \capstart
    \centering 
    \includegraphics[width=\linewidth]{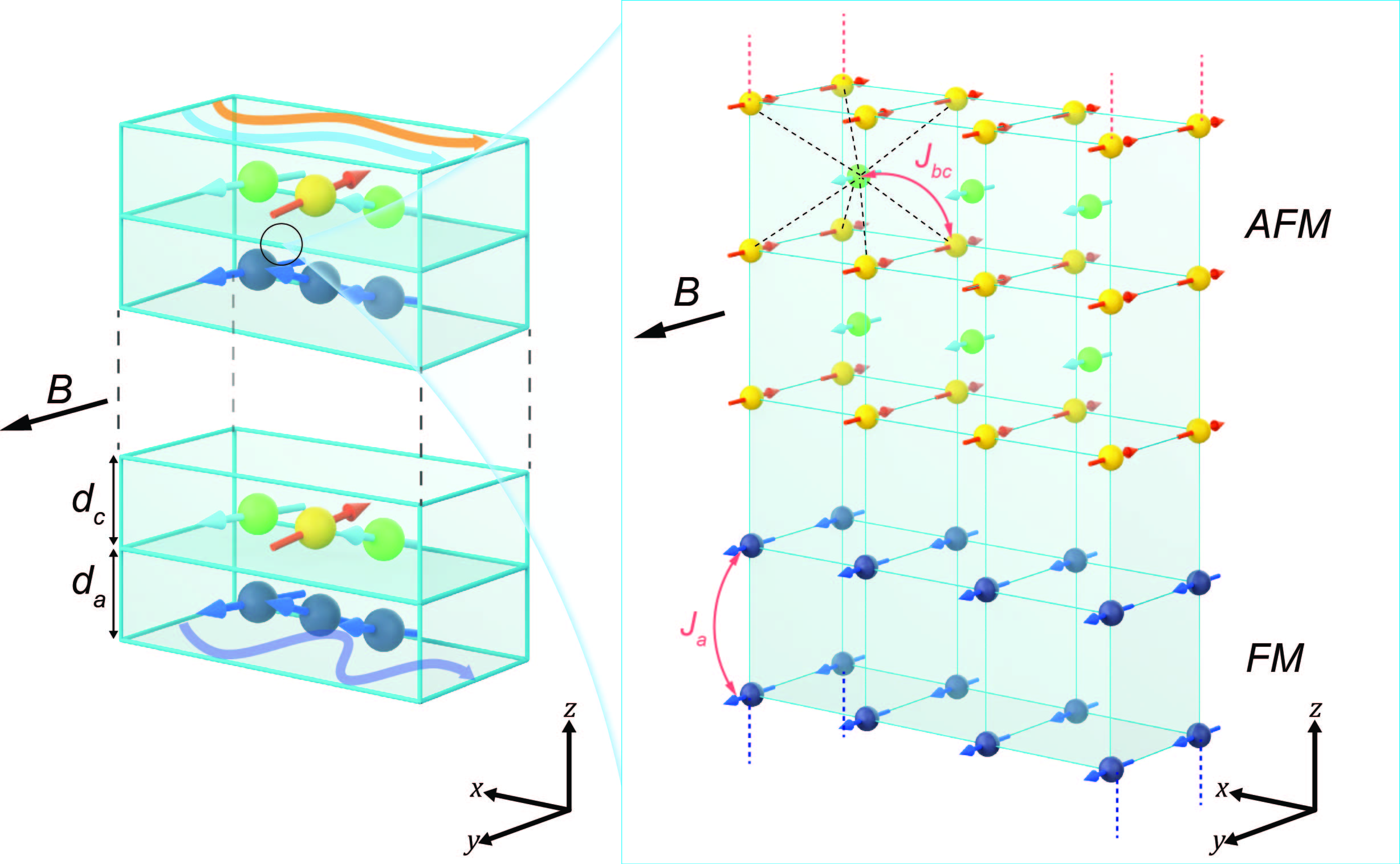} 
    \caption{\label{fig:4}Model of AFM/FM multilayer. The additional cyan arrows are spin moments of atom \(b\) in the AFM layer, pointing in the $+ y$ direction. The AFM layer with the thickness of $d_{c}$ are simple cubic lattice (lattice constant $a_0$) with basis $(1/2,1/2,1/2)a_0$ and Heisenberg exchange constant $J_{bc}$ between nearest type-different atoms.}
    \end{figure}

\label{sec:Model II}    
    \textit{AFM/FM multilayer.---} 
    To demonstrate the versatility of the dipolar-interaction-induced SSH model, we investigate the AFM/FM multilayer structure in Fig. \ref{fig:4}. There are 3 types of spin, \added[id=YF]{with \(a\) in FM layers and \(b\) in AFM layers pointing to $+y$ direction and \(c\) in AFM layers pointing to $-y$ direction.} For a bilayer unit cell, we can write the Hamiltonian as:
    \begin{equation}
      \hat{\mathcal H}_{\mathrm{AF}}
      =\hat{\mathcal H}_{a\mathrm{0}}+\hat{\mathcal H}_{bc\mathrm{0}}+\sum_{l=a,b,c}\hat{\mathcal H}_{l\mathrm{D0}}
      +\sum_{l\neq l'}\hat{\mathcal H}_{ll'\mathrm{D}},
    \end{equation}
     where $\hat{\mathcal H}_{a\mathrm{0}}$ and $\hat{\mathcal H}_{bc\mathrm{0}}$ includes the intralayer exchange interaction and Zeeman interaction, while $\hat{\mathcal H}_{l\mathrm{D0}}$ and $\hat{\mathcal H}_{ll'\mathrm{D}}$ represent dipolar interaction among same and different type of spins. The detail derivations are shown in Sec. C of the Supplemental Material \cite{SM}. After adopting periodic boundary condition in $z$ direction on layers, we arrive at a 6 by 6 Hamiltonian of the multilayer system $\mathbf{H}_{\mathrm{AF}}$.   

    \FloatBarrier 
    \begin{figure}[t] 
    \capstart
    \centering 
    \includegraphics[width=\linewidth]{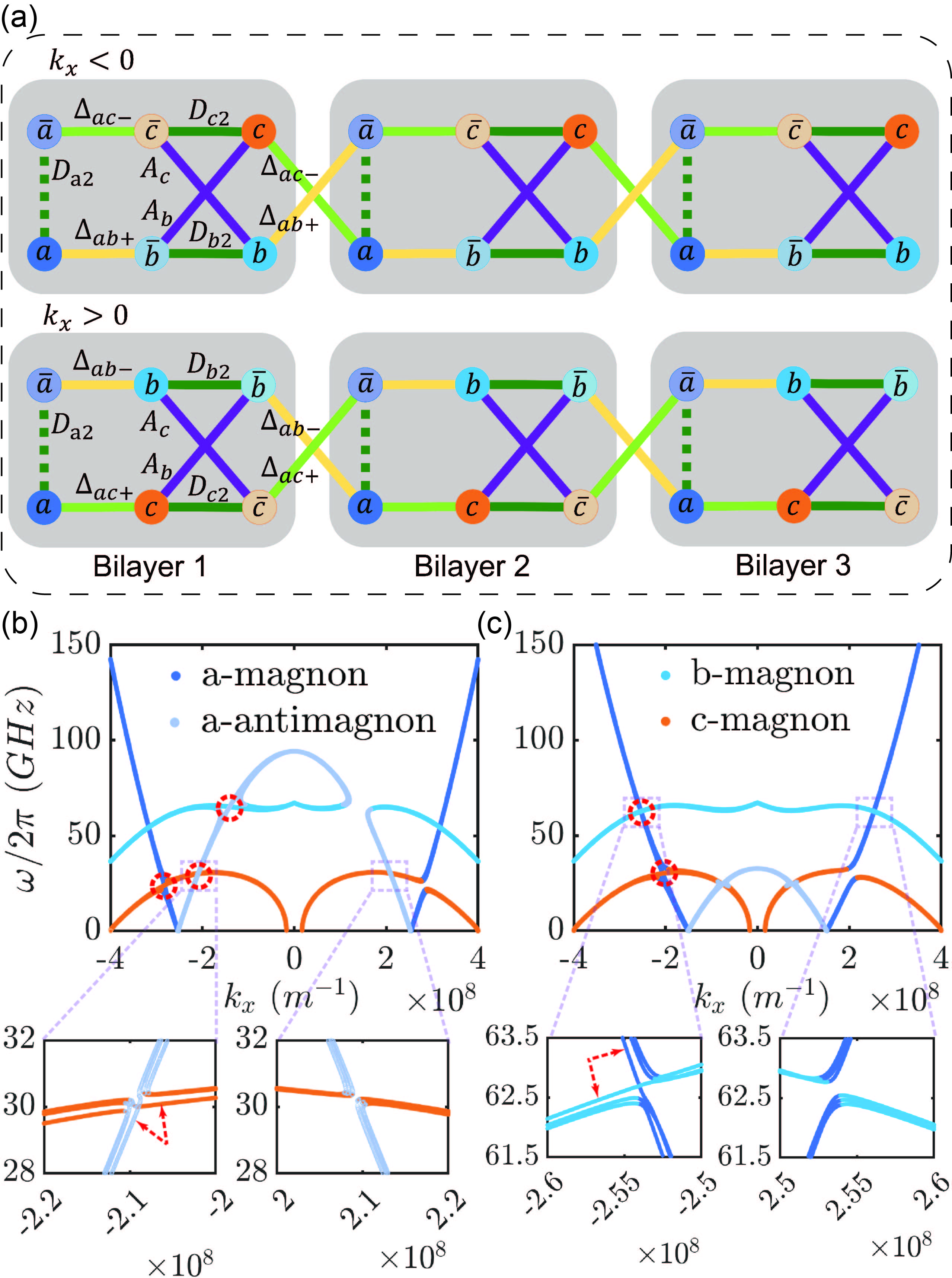} 
    \caption{\label{fig:5}(a)Illustration of two coupled 2D-SSH4 chians for AFM/ FM multilayer. Magnon ($a, b, c$) and antimagnon ($\bar{a}, \bar{b}, \bar{c}$) states are coupled via interlayer (light green, $\Delta_{\pm}$; gold, $\Delta'_{\pm}$), intralayer (dark green, $D_2$) dipolar interactions, and AFM exchange interactions (purple, \(a\)). (b)-(c) Band Structure of 3 AFM/FM bilayer units at different external fields. The overall effective external field for each layers are (b) $B_a^y/\mu_0 = -1\times 10^6 \mathrm{A/m}$ and $B_{b/c}^y/\mu_0 = 5\times 10^5 \mathrm{A/m}$ (nontrivial) (c) $B_a^y/\mu_0 = -2.75\times 10^6 \mathrm{A/m}$ and $B_{b/c}^y/\mu_0 = 5\times 10^5 \mathrm{A/m}$ (nontrivial). Red circles and red arrows indicate nonreciprocal topological surface states}
    \end{figure}
    
    The interaction causing the topological surface state in Fig. \ref{fig:5}(a) is visualized within 3 bilayer unit cells. It is formed by coupling the 2D-SSH4 chain from Fig. \ref{fig:2}(a) and Fig. \ref{fig:3}(a) via AFM exchange interaction.  Similarly, the topological surface states emerge when intrinsic bands of coupled chain's head (Bilayer 1 \(a\) and $\bar{a}$) cross the ones of chain's tails (Bilayer 3 \(b\), $\bar{b}$, \(c\) and $\bar{c}$).

     We calculate the Hamiltonian of a multilayer system consisting of three bilayer unit cells under open boundary conditions. The resulting band structures for two specific cases are shown in Fig. \ref{fig:5}(b)-(c), where the topological surface states are marked by red circles and arrows. The $k_x$ dependence of the dipolar interaction strength causes variations in the relative magnitude at the band crossing points, leading to different numbers of surface states. Consequently, the coupled 2D-SSH4 chains provide a highly tunable platform for realizing topological surface states. 
     \added[id=YF]{Meanwhile, the extra band crossings and three distinct types of atoms create a landscape of highly tunable coupling and chirality states; subpanels (b) and (c) of Fig.~\ref{fig:5} illustrate the dissipative and coherent bulk-coupling regimes, respectively.} 
     In summary, the 2D-SSH4 chain offers a simple and physically intuitive description of complex magnetic multilayer systems.

\label{sec:Conclusion}
    \textit{Conclusion.---} 
    In conclusion, we propose the 2D-SSH4 chain model, which captures the dipolar interaction between antimagnon and magnon and the enlarged Hamiltonian. It is able to describe the topological surface states in FM multilayer, \added[id=YF]{which has the same origin as the classic MSSW.} Through dipolar interaction, we can control spin wave chirality and coherent and dissipative coupling.  
    Finally, we show that our model offers a simple but physically intuitive picture to describe more complex magnetic multilayer systems. \added[id=YF]{Altogether, our model presents a versatile method for exploring and controlling topological magnonic states, spin wave chirality, and coupling in various magnetic multilayers.}

\section*{Acknowledgements} \label{sec:acknowledgements} 
    We thank L. Liu, T. Li, J. Liu, S. Tian, F. Nugraha and R. Liu for helpful discussion. We acknowledge support from Research Grants Council General Research Fund (Nos. 16309924 and 16303322).
\bibliography{arxiv}  

\clearpage
\onecolumngrid
\subsection*{\Large Supplementary Materials}
\normalsize

\let\addcontentsline\oldaddcontentsline  

\setcounter{equation}{0}
\setcounter{figure}{0}
\setcounter{table}{0}
\setcounter{page}{1}
\makeatletter
\renewcommand{\theequation}{S\arabic{equation}}
\renewcommand{\thefigure}{S\arabic{figure}}
\renewcommand{\thetable}{S\arabic{table}}
\renewcommand{\refname}{}

\tableofcontents

\newpage
\section*{Section A: Calculation of Band Structure of FM Multilayer and Chern Number} \label{sec:Appendix_A}
    In this section, we begin by calculating the magnonic band structures of FM multilayer in both parallel and antiparallel configurations, as shown in Figs.\ref{fig:2} and \ref{fig:3}. We then analyze the underlying symmetries of the spin-wave Hamiltonian. Finally, we calculate the associated Chern numbers to characterize the system’s topological properties.
    
    The Hamiltonian of FM bilayer model can be expressed as:
    \begin{equation}
    \label{eq:H_total_appendix}
    \hat{\mathcal H}_{\mathrm{FF}}
       = \sum_{l=a,c}
           \Bigl(
             \hat{\mathcal H}_{l\mathrm{0}}
             + \hat{\mathcal H}_{l\mathrm{D0}}
           \Bigr)
         + \hat{\mathcal H}_{ac\mathrm{D}},
    \end{equation}  
    with
\begin{align}
\label{eq:FFHl0}
\hat{\mathcal H}_{l\mathrm{0}} &=
  -g\mu_{\mathrm B}\sum_{i}\,
     \hat{\bm{S}}_{l i}\!\cdot\!\bm{B}_{l}
  +J_{l}\!\!\sum_{\langle i,j\rangle}
     \hat{\bm{S}}_{l i}\!\cdot\!\hat{\bm{S}}_{l j},
\end{align}
\begin{align}
\hat{\mathcal H}_{l\mathrm D0} &=
  \frac{\mu_{0}\bigl(g\mu_{\mathrm B}\bigr)^{2}}{2}
  \sum_{i\neq j}
    \frac{
      R_{ij}^{2}\,\hat{\bm{S}}_{l i}\!\cdot\!\hat{\bm{S}}_{l j}
      -3\bigl(\bm{R}_{ij}\!\cdot\!\hat{\bm{S}}_{l i}\bigr)
           \bigl(\bm{R}_{ij}\!\cdot\!\hat{\bm{S}}_{l j}\bigr)}
         {R_{ij}^{5}},
\end{align}
\begin{align}
\hat{\mathcal H}_{ac\mathrm D} &=
  \frac{\mu_{0}\bigl(g\mu_{\mathrm B}\bigr)^{2}}{2}
  \sum_{i\in a}\sum_{j\in c}
    \frac{
      R_{ij}^{2}\,\hat{\bm{S}}_{a i}\!\cdot\!\hat{\bm{S}}_{c j}
      -3\bigl(\bm{R}_{ij}\!\cdot\!\hat{\bm{S}}_{a i}\bigr)
           \bigl(\bm{R}_{ij}\!\cdot\!\hat{\bm{S}}_{c j}\bigr)}
         {R_{ij}^{5}}.
\end{align}
    where $l=a$ represents the lower FM layer with magnetization pointing $+y$ direction and $l=c$ represents the upper FM layer with magnetization pointing $-y$ direction. $\hat{\mathcal H}_{l\mathrm{0}}$ represents the Hamiltonian of Zeeman interaction and intralayer exchange interaction of layer $l$. $J_l<0$ for FM intralayer exchange coupling. $\hat{\mathcal H}_{l\mathrm D0}$ is the Hamiltonian of intralayer dipolar interaction of layer $l$. $\bm{R}_{ij}=\bm{r}_{j}-\bm{r}_{i}$ is the displacement from spin $i$ to spin $j$. And $\hat{\mathcal H}_{ac\mathrm D}$ is the Hamiltonian of the dipolar interlayer interaction of both layers.
    
    Assuming that deviations around the direction of magnetization are small, we apply the Holstein–Primako (HP) transformation \cite{dipolar_multilayer_2_AFM, HP_Trans}:
\begin{align}
\hat{S}^{y}_{a i} &= S_{a} - \hat{a}^{\dagger}_{i}\hat{a}_{i}, \label{eq:HP_Say}\\
\hat{S}^{+}_{a i} &= \hat{S}^{z}_{a i}+i\hat{S}^{x}_{a i}
                 = \sqrt{2S_{a}-\hat{a}^{\dagger}_{i}\hat{a}_{i}}\;\hat{a}_{i}
                 \simeq \sqrt{2S_{a}}\;\hat{a}_{i}, \label{eq:HP_Sap}\\
\hat{S}^{-}_{a i} &= \hat{S}^{z}_{a i}-i\hat{S}^{x}_{a i}
                 = \hat{a}^{\dagger}_{i}\sqrt{\,2S_{a}-\hat{a}^{\dagger}_{i}\hat{a}_{i}}
                 \simeq \sqrt{2S_{a}}\;\hat{a}^{\dagger}_{i}, \label{eq:HP_Sam}
\end{align}

\begin{align}
\hat{S}^{y}_{c j} &= -S_{c} + \hat{c}^{\dagger}_{j}\hat{c}_{j}, \label{eq:HP_Scy}\\
\hat{S}^{+}_{c j} &= \hat{S}^{z}_{c j}+i\hat{S}^{x}_{c j}
                 = \hat{c}^{\dagger}_{j}\sqrt{\,2S_{c}-\hat{c}^{\dagger}_{j}\hat{c}_{j}}
                 \simeq \sqrt{2S_{c}}\;\hat{c}^{\dagger}_{j}, \label{eq:HP_Scp}\\
\hat{S}^{-}_{c j} &= \hat{S}^{z}_{c j}-i\hat{S}^{x}_{c j}
                 = \sqrt{2S_{c}-\hat{c}^{\dagger}_{j}\hat{c}_{j}}\;\hat{c}_{j}
                 \simeq \sqrt{2S_{c}}\;\hat{c}_{j}. \label{eq:HP_Scm}
\end{align}
    where $\hat{a}(\hat{c})_{i}^{\dagger}$ represents the creation operator of a magnon in site $i$ of layer $a$($c$). 

    Plug HP transformation into ~\eqref{eq:FFHl0} and only keep the quadratic terms, we have:
\begin{equation}
\begin{aligned}
\sum_{l=a,c}\hat{\mathcal H}_{l\mathrm{0}} &=
  -g\mu_{\mathrm B}\sum_{i}
    \bigl(
      S_{a}B_{a}^{y}-B_{a}^{y}\,\hat{a}_{i}^{\dagger}\hat{a}_{i}
      -S_{c}B_{c}^{y}+B_{c}^{y}\,\hat{c}_{i}^{\dagger}\hat{c}_{i}
    \bigr) \\[2pt]
 &\quad+ \frac{1}{6}\sum_{i}\sum_{\bm{\delta}} J_{a}
    \bigl\{
      S_{a}^{2}
      +S_{a}\bigl(
         -\hat{a}_{i}^{\dagger}\hat{a}_{i}
         -\hat{a}_{i+\bm{\delta}}^{\dagger}\hat{a}_{i+\bm{\delta}}
         +\hat{a}_{i}\hat{a}_{i+\bm{\delta}}^{\dagger}
         +\hat{a}_{i}^{\dagger}\hat{a}_{i+\bm{\delta}}
      \bigr)
    \bigr\} \\[2pt]
 &\quad+ \frac{1}{6}\sum_{i}\sum_{\bm{\delta}} J_{c}
    \bigl\{
      S_{c}^{2}
      +S_{c}\bigl(
         -\hat{c}_{i}^{\dagger}\hat{c}_{i}
         -\hat{c}_{i+\bm{\delta}}^{\dagger}\hat{c}_{i+\bm{\delta}}
         +\hat{c}_{i}\hat{c}_{i+\bm{\delta}}^{\dagger}
         +\hat{c}_{i}^{\dagger}\hat{c}_{i+\bm{\delta}}
      \bigr)
    \bigr\}.
\end{aligned}
\end{equation}
    where $\bm{\delta}$ denotes the displacement vectors that connect nearest-neighbour sites, and the site index $i$ is a shorthand for the lattice position $\bm r_{i}$. 
    
    The Fourier Transform of creation and annihilation operators are:
\begin{equation}
\hat{a}_{i} =
  \frac{1}{\sqrt{N_{a}}}\sum_{\bm{k}\in\text{BZ}}
    e^{-i\bm{k}\cdot\bm{r}_{i}}\,
    \hat{a}_{\bm{k}},\qquad
\hat{c}_{i} =
  \frac{1}{\sqrt{N_{c}}}\sum_{\bm{k}\in\text{BZ}}
    e^{-i\bm{k}\cdot\bm{r}_{i}}\,
    \hat{c}_{\bm{k}},
\end{equation}

\begin{equation}
\hat{a}^{\dagger}_{i} =
  \frac{1}{\sqrt{N_{a}}}\sum_{\bm{k}\in\text{BZ}}
    e^{\,i\bm{k}\cdot\bm{r}_{i}}\,
    \hat{a}^{\dagger}_{\bm{k}},\qquad
\hat{c}^{\dagger}_{i} =
  \frac{1}{\sqrt{N_{c}}}\sum_{\bm{k}\in\text{BZ}}
    e^{\,i\bm{k}\cdot\bm{r}_{i}}\,
    \hat{c}^{\dagger}_{\bm{k}} .
\end{equation}

    Plug them into the Hamiltonian and utilize the relation:
\begin{equation}
\bigl[\,\hat{a}_{\bm{k}},\hat{a}^{\dagger}_{\bm{k}'}\bigr]
  =\delta_{\bm{k}\bm{k}'},\qquad
\bigl[\,\hat{c}_{\bm{k}},\hat{c}^{\dagger}_{\bm{k}'}\bigr]
  =\delta_{\bm{k}\bm{k}'} ,
\end{equation}

\begin{equation}
\sum_{j} e^{i(\bm{k}-\bm{k}')\cdot\bm{r}_{j}}
  = N\,\delta_{\bm{k}\bm{k}'}.
\end{equation}
We arrive at:
\begin{equation}
\label{eq:H0_k}
\begin{aligned}
\sum_{l=a,c}\hat{\mathcal H}_{l0} &=
  -g\mu_{\mathrm B}\Bigl(
       S_{a}B_{a}^{y}N_{a}
      -S_{c}B_{c}^{y}N_{c}
      +\sum_{\bm{k}}
        \bigl(
          -B_{a}^{y}\,\hat{a}^{\dagger}_{\bm{k}}\hat{a}_{\bm{k}}
          +B_{c}^{y}\,\hat{c}^{\dagger}_{\bm{k}}\hat{c}_{\bm{k}}
        \bigr)
    \Bigr) \\[4pt]
 &\quad+ \sum_{l=a,c} S_{l}^{2} J_{l} N_{l}
        -\sum_{l=a,c}\sum_{\bm{k}}
           4S_{l}J_{l}\,
           \hat{l}^{\dagger}_{\bm{k}}\hat{l}_{\bm{k}}
           \sum_{i=x,y,z}\sin^{2}\!\bigl(\tfrac{a_{0}k_{i}}{2}\bigr),
\end{aligned}
\end{equation}
Assume the continuous magnetization, replace $\sum_{\bm{k}}$ with $\int \frac{d^3k}{(2\pi)^3}$:
    \begin{equation}
\label{eq:H0_k_cont}
\begin{aligned}
\sum_{l=a,c}\hat{\mathcal H}_{l0} &=
  -g\mu_{\mathrm B}\Bigl(
       S_{a}B_{a}^{y}N_{a}
      -S_{c}B_{c}^{y}N_{c}
      +\int\!\frac{d^{3}k}{(2\pi)^{3}}
        \bigl(
          -B_{a}^{y}\,\hat{a}^{\dagger}_{\bm{k}}\hat{a}_{\bm{k}}
          +B_{c}^{y}\,\hat{c}^{\dagger}_{\bm{k}}\hat{c}_{\bm{k}}
        \bigr)
    \Bigr) \\[4pt]
 &\quad+ \sum_{l=a,c} S_{l}^{2} J_{l} N_{l}
        -\sum_{l=a,c}\int\!\frac{d^{3}k}{(2\pi)^{3}}
           4S_{l}J_{l}\,
           \hat{l}^{\dagger}_{\bm{k}}\hat{l}_{\bm{k}}
           \sum_{i=x,y,z}\sin^{2}\!\bigl(\tfrac{a_{0}k_{i}}{2}\bigr).
\end{aligned}
\end{equation}
    
    For $\hat{H}_{acD}$ and $\hat{H}_{lD0}$, we follow the integration approximation in ref\cite{SSH1_Luqiao}:
    Rewrite $\hat{H}_{acD}$ into matrix form: 
    \begin{equation}
    \hat{\mathcal H}_{ac\mathrm D} =
      \frac{\mu_{0}\bigl(g\mu_{\mathrm B}\bigr)^{2}}{4}
      \sum_{i\in a}\sum_{j\in c}
      \hat{\bm S}_{c}^{\mathsf T}\!\bigl(\bm r_{j}\bigr)\,
      \bm{\mathbf F}\bigl(\bm r_{i}-\bm r_{j}\bigr)\,
      \hat{\bm S}_{a}\!\bigl(\bm r_{i}\bigr)
      \;+\;\text{H.c.}
    \end{equation}
    where 
    \begin{equation}
    \vspace{0.1cm}
    \mathbf{F}(\bm{r}_i-\bm{r}_j) = 
    \frac{1}{R_{ij}^{2}}\begin{bmatrix}
      3X_{ij}^{2}-R_{ij}^{2}&  3X_{ij}Y_{ij}& 3X_{ij}Z_{ij}\\
      3X_{ij}Y_{ij}&  3Y_{ij}^{2}-R_{ij}^{2}& 3Y_{ij}Z_{ij}\\
      3X_{ij}Z_{ij}&  3Y_{ij}Z_{ij}&  3Z_{ij}^{2}-R_{ij}^{2}
    \end{bmatrix},
    \end{equation}
    with 
    \[
    \bm R_{ij}=
    \begin{bmatrix}
      X_{ij}\\[2pt] Y_{ij}\\[2pt] Z_{ij}
    \end{bmatrix},\qquad
    R_{ij}=|\bm R_{ij}|.
    \]

    Replace the summation with integration, assuming continuous magnetization in each layer, and apply the Fourier Transform, for a simple cubic lattice:
    \begin{equation}
    \hat{\mathcal H}_{ac\mathrm D}=
    \frac{\mu_{0}\bigl(g\mu_{\mathrm B}\bigr)^{2}}{4}
    \int\!\frac{\mathrm d^{3}k}{(2\pi)^{3}}\;
    \hat{\bm S}_{c}^{\mathsf T}(\bm k)\,
    \mathbf F(\bm k)\,
    \hat{\bm S}_{a}(\bm k)
    +\text{H.c.}
    \end{equation}
    where 
    \begin{equation}
    \mathbf F(\bm k)=\frac{4\pi}{3k^{2}}
    \begin{bmatrix}
    3k_{x}^{2}-k^{2} & 3k_{x}k_{y} & 3k_{x}k_{z}\\
    3k_{x}k_{y} & 3k_{y}^{2}-k^{2} & 3k_{y}k_{z}\\
    3k_{x}k_{z} & 3k_{y}k_{z} & 3k_{z}^{2}-k^{2}
    \end{bmatrix},
    \qquad k\equiv|\bm k|.
    \end{equation}
    If $k$ is not large compared with the first Brillouin zone (BZ).
    
    Observe that: 
    \begin{equation}
    \hat{\bm S}_{a(c)}(\bm r)=
    \hat{\bm S}_{a(c)}(\bm r_{\parallel})\,
    \Theta_{a(c)}(z;d_{a(c)}),
    \end{equation}
    where 
    \begin{equation}
    \Theta_{c}(z;d_{c})=
    \begin{cases}
    1,& 0<z<d_{c},\\
    0,& \text{otherwise},
    \end{cases}
    \qquad
    \Theta_{a}(z;d_{a})=
    \begin{cases}
    1,& -d_{a}<z<0,\\
    0,& \text{otherwise}.
    \end{cases}
    \end{equation}
    Define 
    \[
\bm k_{\parallel}\equiv (k_{x},k_{y})^{\mathsf T},\qquad
k_{\parallel}=|\bm k_{\parallel}|=\sqrt{k_{x}^{2}+k_{y}^{2}}.
\]

    Thus,
    \begin{equation}
    \hat{\mathcal H}_{ac\mathrm D}=
    \frac{\mu_{0}\bigl(g\mu_{\mathrm B}\bigr)^{2}}{4}
    \int\!\frac{\mathrm d^{2}k_{\parallel}}{(2\pi)^{2}}\;
    \hat{\bm S}_{c}^{\mathsf T}(\bm k_{\parallel})\,
    \mathbf G(\bm k_{\parallel})\,
    \hat{\bm S}_{a}(\bm k_{\parallel})
    +\text{H.c.}
    \end{equation}
    where 
    \begin{equation}
    \begin{aligned}
    \mathbf G(\bm k_{\parallel})
      &=\int_{-\infty}^{+\infty}\!\frac{\mathrm d k_{z}}{2\pi}\,
        \tilde\Theta_{c}(k_{z};d_{c})\,
        \mathbf F(\bm k)\,
        \tilde\Theta_{a}(k_{z};d_{a}) \\[4pt]
      &=\frac{4\pi}{2k_{\parallel}^{3}}\,
        \bigl(1-e^{-k_{\parallel}d_{a}}\bigr)\,
        \bigl(1-e^{-k_{\parallel}d_{c}}\bigr)
        \begin{bmatrix}
          -k_{x}^{2} & -k_{x}k_{y} & i\,k_{x}k_{\parallel}\\
          -k_{x}k_{y} & -k_{y}^{2} & i\,k_{y}k_{\parallel}\\
           i\,k_{x}k_{\parallel} & i\,k_{y}k_{\parallel} & k_{\parallel}^{2}
        \end{bmatrix}.
    \end{aligned}
\end{equation}

    To simplify the case, we consider spin waves that only propagate in the $x$ direction, i.e. $k_{\parallel}=k_x,\ k_y=0$. Then we apply the HP transformation, and we arrive: 
    \begin{equation}
\label{eq:H_acD_kx}
\hat{\mathcal H}_{ac\mathrm D} = \frac{1}{2}
\mu_{0}\bigl(g\mu_{\mathrm B}\bigr)^{2}\,
\pi\sqrt{S_{a}S_{c}}
\int\!\frac{\mathrm d k_{x}}{2\pi}\;
\frac{\bigl[1-e^{-k_{x}d_{a}}\bigr]\bigl[1-e^{-k_{x}d_{c}}\bigr]}{k_{x}^{2}}
\bigl[
  (\hat c_{k_{x}}^{\dagger}\hat a_{k_{x}})(|k_{x}|-k_{x})
 +(\hat c_{k_{x}}\hat a_{k_{x}}^{\dagger})(|k_{x}|+k_{x})
\bigr]
+\text{H.c.}
\end{equation}
    Finally, apply a similar treatment to the Hamiltonian of intralayer dipolar interaction:
\begin{equation}
\begin{aligned}
\sum_{l=a,c}\hat{\mathcal H}_{l\mathrm D0}
  &=\frac{\mu_{0}\bigl(g\mu_{\mathrm B}\bigr)^{2}}{4}
    \sum_{l=a,c}\!
    \int\!\frac{\mathrm d^{3}k}{(2\pi)^{3}}\;
      \hat{\bm S}_{l}^{\mathsf T}(\bm k)\,
      \mathbf F_{l}(\bm k)\,
      \hat{\bm S}_{l}(\bm k)
    +\text{H.c.} \\[4pt]
  &=\frac{\mu_{0}\bigl(g\mu_{\mathrm B}\bigr)^{2}}{4}
    \sum_{l=a,c}\!
    \int\!\frac{\mathrm d^{2}k_{\parallel}}{(2\pi)^{2}}\;
      \hat{\bm S}_{l}^{\mathsf T}(\bm k_{\parallel})\,
      \mathbf G_{l}(\bm k_{\parallel})\,
      \hat{\bm S}_{l}(\bm k_{\parallel})
    +\text{H.c.}
\end{aligned}
\end{equation}
    where 
\begin{equation}
\mathbf G_{l}(\bm k_{\parallel}) =
\frac{4\pi d_{l}}{k_{\parallel}^{2}}
\begin{bmatrix}
k_{x}^{2}\bigl(1-\Phi_{l}\bigr) & k_{x}k_{y}\bigl(1-\Phi_{l}\bigr) & 0 \\
k_{x}k_{y}\bigl(1-\Phi_{l}\bigr) & k_{y}^{2}\bigl(1-\Phi_{l}\bigr) & 0 \\
0 & 0 & k_{\parallel}^{2}\Phi_{l}
\end{bmatrix},
\qquad
\Phi_{l}=\frac{1-e^{-|k_{\parallel}|d_{l}}}{|k_{\parallel}|d_{l}}.
\end{equation}
    
    Again assume $k_{\parallel}=k_x,\ k_y=0$, and only keep the quadratic terms, we have:
\begin{equation}
\label{eq:H_lD0_kx}
\sum_{l=a,c}\hat{\mathcal H}_{l\mathrm D0}=
\frac{\pi\mu_{0}\bigl(g\mu_{\mathrm B}\bigr)^{2}}{2}
\sum_{l=a,c} d_{l}S_{l}
\int\!\frac{\mathrm d k_{x}}{2\pi}\;
\Bigl[
  \bigl(\hat l_{k_{x}}^{\dagger}\hat l_{k_{x}}
       +\hat l_{k_{x}}\hat l_{k_{x}}^{\dagger}\bigr)
 +\bigl(\hat l_{k_{x}}^{\dagger}\hat l_{k_{x}}^{\dagger}
       +\hat l_{k_{x}}\hat l_{k_{x}}\bigr)\,(2\Phi_{l}-1)
\Bigr]
+\text{H.c.}
\end{equation}
    By defining
\begin{equation}
\label{eq:Nambu_basis_FF}
\hat{\bm{\beta}}_{k_{x}}\equiv
\bigl[\,
  \hat a_{k_{x}}\;,
  \hat c_{k_{x}}\;,
  \hat a_{-k_{x}}^{\dagger}\;,
  \hat c_{-k_{x}}^{\dagger}
\bigr]^{\mathsf T},
\qquad
\end{equation}
we can write the total Hamiltonian into matrix form:
\begin{equation}
\hat{\mathcal H}_{\mathrm{FF}}=
\frac{1}{2}\sum_{k_{x}}
\hat{\bm{\beta}}_{k_{x}}^{\dagger}\,
\mathbf H_{\mathrm{BdG}}^{\mathrm{FF}}\,
\hat{\bm{\beta}}_{k_{x}},
\end{equation}
where
\begin{equation}
\mathbf H_{\mathrm{BdG}}^{\mathrm{FF}}=
\begin{bmatrix}
\omega_{a}+D_{a1} & \Delta_{+} & D_{a2} & 0 \\
\Delta_{+} & \omega_{c}+D_{c1} & 0 & D_{c2} \\
D_{a2} & 0 & \omega_{a}+D_{a1} & \Delta_{-} \\
0 & D_{c2} &
\Delta_{-} & \omega_{c}+D_{c1}
\end{bmatrix},
\end{equation}
    with
\begin{align}
\omega_{l}      &= -S_{l}J_{l}a_{0}^{2}k_{x}^{2}
                  \; \pm\; g\mu_{\mathrm B}B_{l}^{y},
                  (l=a:+,\;l=c:-) \\[4pt]
\Delta_{\pm}    &= \pi\mu_{0}\bigl(g\mu_{\mathrm B}\bigr)^{2}
                  \sqrt{S_{a}S_{c}}\,
                  \Phi_{a}\Phi_{c}\,d_{a}d_{c}\,
                  \bigl(|k_{x}|\pm k_{x}\bigr), \\[4pt]
D_{l1}          &= \pi\mu_{0}\bigl(g\mu_{\mathrm B}\bigr)^{2}\,d_{l}S_{l}, \\[4pt]
D_{l2}          &= \pi\mu_{0}\bigl(g\mu_{\mathrm B}\bigr)^{2}\,d_{l}S_{l}
                  \bigl(-2\Phi_{l}+1\bigr). 
\end{align}
    Note that both $\mathbf H_{\mathrm{BdG}}^{\mathrm{FF}}$ and
    $\hat{\mathcal H}_{\mathrm{FF}}$ are Hermitian, consistent with the basic axiom of quantum mechanics.  
    To diagonalize the bosonic system while preserving the commutation
    relation, we apply the generalized–eigenvalue approach
    \cite{Basic_Review}: We solve for the eigenvalues of
    $\bm {\eta}\,\mathbf H_{\mathrm{BdG}}^{\mathrm{FF}}$, where
    \[
    \bm{\eta}\equiv\mathrm{diag}\,(1,1,-1,-1).
    \]
    The resulting spectrum corresponds to the classical Hamiltonian
    that can be obtained from an effective field coupled to the LLG
    equations.  
    Quantum and classical descriptions meet when we retain only the
    quadratic (bilinear) ladder–operator terms in the quantum Hamiltonian.
    Although $\bm{\eta}\,\mathbf H_{\mathrm{BdG}}^{\mathrm{FF}}$ is
    generally non-Hermitian, the underlying  physical Hamiltonian
    remains Hermitian; the apparent non-Hermiticity arises solely from the doubling of bosonic degrees of freedom required to encode the
    particle–hole–like Nambu structure.

    In addition, if layer $c$ is positioned below layer $a$ while all
other parameters remain unchanged, the Nambu basis keeps the same form as ~\eqref{eq:Nambu_basis_FF}
but the Bogoliubov–de Gennes (BdG) matrix satisfies:
\begin{equation}
\mathbf H_{\mathrm{BdG}}^{\mathrm{FF,reverse}}(k_{x})=
\mathbf H_{\mathrm{BdG}}^{\mathrm{FF}}(-k_{x}).
\end{equation}
    
    For spins in both layer $a$ and $c$ align in the same $+y$ direction, we only modify the HP transformation of layer $c$ to the exact form as layer $a$, and we will arrive at
    \begin{equation}
    \mathbf H_{\mathrm{BdG}}^{\mathrm{FF}\prime}=\begin{bmatrix}
    \omega_a'+D_{a1}&  0&  D_{a2}& \Delta_+'\\
    0&  \omega_c'+D_{c1}&  \Delta_-'& D_{c2}\\
    D_{a2}&  \Delta_-'&  \omega_a'+D_{a1}& 0\\
    \Delta_+'&  D_{c2}&  0& \omega_c'+D_{c1}
    \end{bmatrix},
    \end{equation} 
    with only modification on:
    \begin{align}
    \omega_l' &= -S_{l}J_{l}a_0^2k_x^2 \ + \ g\mu_BB_{l}^y, \\[4pt]
    \Delta_{\pm }' &= \pi\mu_0(g\mu_B)^2\sqrt{S_aS_c}\Phi_a\Phi_cd_ad_c(-|k_x|\pm k_x). 
    \end{align}

    Within the generalized–eigenvalue framework
    $\bm{\eta}\mathbf H_{\mathrm{BdG}}\bm{\psi}=\omega\bm{\psi}$,
    the metric
    $\bm{\eta}=\mathrm{diag}(1,1,-1,-1)$
    multiplies the antimagnon sector by $-1$, thereby imparting
    an additional phase of $\pi$ to every magnon–antimagnon matrix element,
    while leaving magnon–magnon and antimagnon–antimagnon terms with no phase difference.
    As a result, the net interlayer coupling may be written
    $g\,e^{i\phi}$ for the net coupling strength $g$, with
    $\phi=0$ for magnon–magnon or antimagnon–antimagnon blocks and $\phi=\pi$ for magnon–antimagnon blocks. 
    To make this phase dependence explicit, we cast the bilayer into the canonical two-mode Hamiltonian
    \begin{equation}
    \hat H
    =\hbar\omega_a \hat a^{\dagger}\hat a
    +\hbar\omega_c \hat c^{\dagger}\hat c
    +\hbar g\!\left(
      \hat a^{\dagger}\hat c
     +e^{i\phi}\hat c^{\dagger}\hat a
    \right),
    \label{eq:phase_H}
    \end{equation}
    where $\hat a^{\dagger}(\hat a)$ and
    $\hat c^{\dagger}(\hat c)$ create (annihilate) magnons in layers $a$ and $c$, respectively.
    For $\phi=0$, Eq.~\eqref{eq:phase_H} yields the familiar level-repulsive anticrossing that defines
   coherent coupling.%
    Conversely, $\phi=\pi$ leads to level attraction and imaginary-frequency states between the anticrossing as the signature of dissipative coupling.
    This phase-based picture unifies the two hybridization regimes reported in cavity-magnon systems  \cite{dissipative_1, dissipative_2, dissipative_3}, and directly explains the level-repulsive versus level-attractive anticrossings that emerge in the band structure of our ferromagnetic bilayer model.

    To construct the SSH like model with 2-dimensional stacking, we stack the bilayer unit cells to a multilayer. It is easy to see that the upmost and lowest layers only couple to the layer next to them when $k_x>0$; however, every layer in between has two layers to couple to when $k_x>0$ or $k_x<0$. Thus, the couple mode exists for any $k_x$. This results in the bulk-edge correspondence and thus the SSH-like model. When $k_x<0$, there will be intrinsic modes in the upmost and lowest layer as surface propagating modes. 

    In each layer, further assuming that spin moments remain uniform along $z$, therefore, no spin wave propagation along $z$ ($k_z=0$). However, in the multilayer system, where each layer acts as a macrospin moment, spin waves do exist along $z$. Supposing we have $M$ bilayer unit cells in total, we can define the magnon state of each layer to be $\left | n, l  \right \rangle$ where $n=1,...,M$ and $l = a,c$. It is noticeable that the total state is the tensor product of this layer state and the Fock state in each layer. We apply the periodic boundary condition along the $z$ direction:
    \begin{equation}
    \left | M+1, l  \right \rangle = \left | 1, l  \right \rangle 
    \end{equation}
    
\begin{figure} 
    \capstart
    \centering 
    \includegraphics[width=0.5\textwidth]{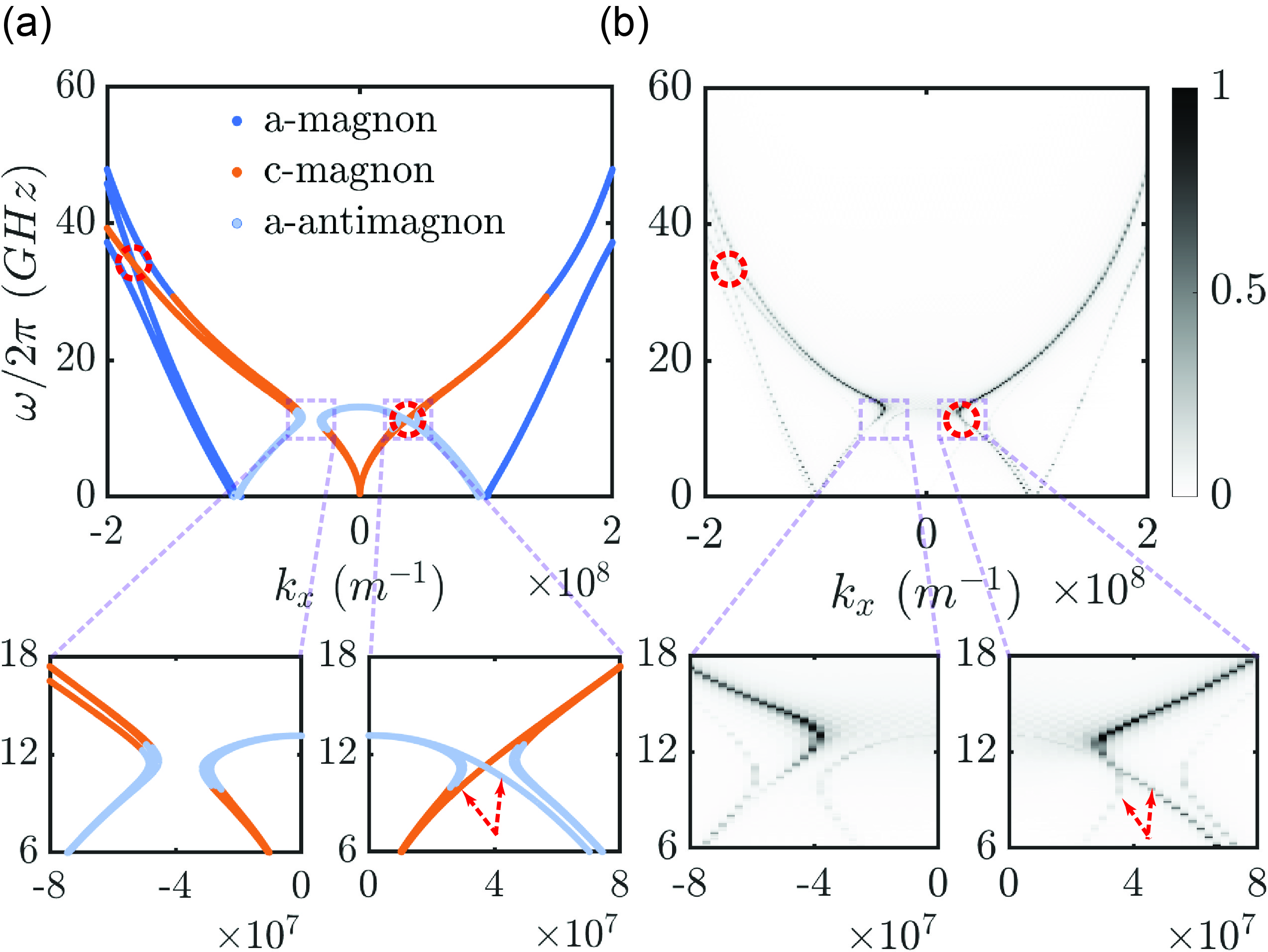} 
    \caption{\label{fig:2.1}(a) Calculated band structure for three antiparallel FM-bilayer unit cells; (b) Simulated counterpart for two unit cells, both obtained under the overall effective external field: $B_a^y/\mu_0 = -4.5\times 10^5 \mathrm{A/m}$ and $B_c^y/\mu_0 = 0 \mathrm{A/m}$ (nontrivial). Red circles and red arrows indicate nonreciprocal topological surface states.}
    \end{figure}

\begin{figure} 
    \capstart
    \centering 
    \includegraphics[width=0.5\textwidth]{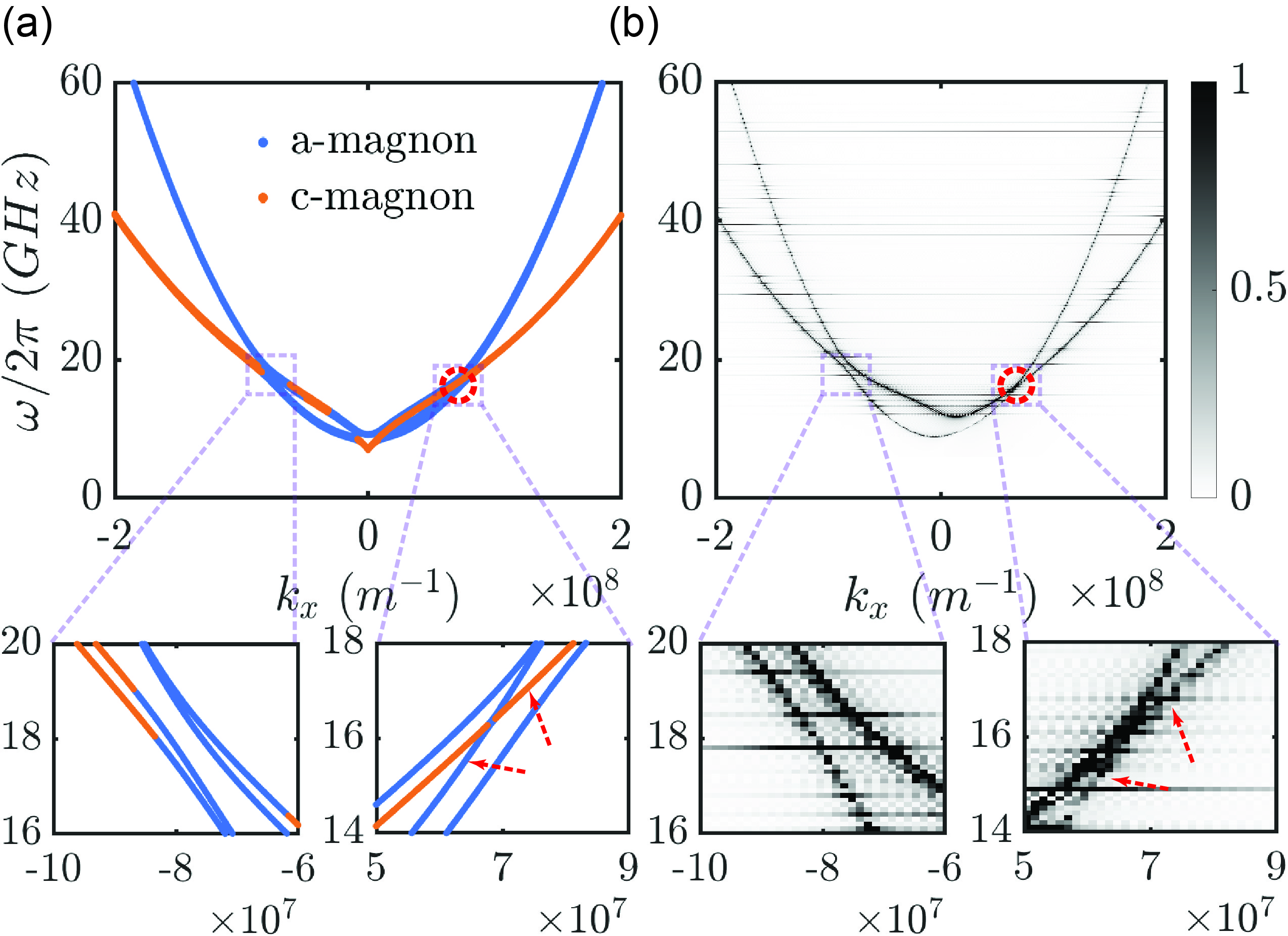} 
    \caption{\label{fig:3.1}(a) Calculated band structure for three parallel FM-bilayer unit cells; (b) Simulated counterpart for one unit cell, both obtained under the overall effective external field: $B_a^y/\mu_0 = 2\times 10^5 \mathrm{A/m}$ and $B_c^y/\mu_0 = 0.5 \mathrm{A/m}$ (nontrivial). Red circles and red arrows indicate nonreciprocal topological surface states.}
    \end{figure}

    Assuming open boundary condition (OBC) in $z$ direction, we can calculate the band structure where surface states occur. Fig. \ref{fig:2.1} and \ref{fig:3.1} are two extra cases of anti-parallel and parallel configuration.
    
    Assuming PBC along $z$ direction, the Fourier Transform for magnon state along $z$ direction gives: 
    \begin{equation}
    \left | n, l  \right \rangle = \frac{1}{\sqrt{M}}\sum_{k_z}^{}e^{-in k_z (d_a+d_c)}\left | k_z, l  \right \rangle 
    \end{equation}
    $k_z$ here means the wave propagating between layers, not within a single layer.

    Define $d=d_a+d_c$. For layer $a$ and $c$ pointing toward opposite directions, adopting the periodic boundary condition, we can get a 4 by 4 Hamiltonian $ \mathbf{H}_{\mathrm{FF}}$. Similarly, for spins in both layer $a$ and $c$ align in the same $+y$ direction, we can arrive at $ \mathbf{H}_{\mathrm{FF}}'$.
    \begin{equation}
    \mathbf{H}_{\mathrm{FF}} = \begin{bmatrix}
    \omega_a+D_{a1}&  \Delta_+ +e^{-ik_zd}\Delta_-&  D_{a2}& 0\\
    \Delta_+ +e^{ik_zd}\Delta_-&  \omega_c+D_{c1}&  0& D_{c2}\\
    D_{a2}&  0&  \omega_a+D_{a1}& \Delta_- +e^{ik_zd}\Delta_+\\
    0&  D_{c2}&  \Delta_-+e^{-ik_zd}\Delta_+& \omega_c+D_{c1}
    \end{bmatrix}
    \end{equation}

    \begin{equation}
    \mathbf{H}_{\mathrm{FF}}'=\begin{bmatrix}
    \omega_a'+D_{a1}&  0&  D_{a2}& \Delta_+' +e^{-ik_zd}\Delta_-' \\
    0&  \omega_c'+D_{c1}&  \Delta_-' +e^{ik_zd}\Delta_+' & D_{c2}\\
    D_{a2}&  \Delta_-' +e^{-ik_zd}\Delta_+' &  \omega_a'+D_{a1}& 0\\
    \Delta_+' +e^{ik_zd}\Delta_-' &  D_{c2}&  0& \omega_c'+D_{c1}
    \end{bmatrix}
    \end{equation} 
   
    In the calculation we assume: $gu_B/\hbar = 1.76 \times 10^{11} \ \mathrm{rad/T}$,
    $S_a\hbar = 7.95 \times 10^{-7}\ \mathrm{J\,s}$, 
    $S_c\hbar = 4.20 \times 10^{-6}\ \mathrm{J\,s}$, 
    $d_a=d_c=10 \ \mathrm{nm}$, 
    $-S_{a}J_{a}a_0^2 = 9.30\times 10^{-6}\ \mathrm{m^2/s}$ and 
    $-S_{b}J_{b}a_0^2 = 4.14\times 10^{-6}\ \mathrm{m^2/s}$.
    
    Both of them obey the particle–hole symmetry (PHS) for a bosonic system:
    \begin{equation}
         \mathbf{H}_{ \bm{k}}=\left(\begin{array}{cc}\mathbf{h}( \bm{k}) & \mathbf{\Delta}( \bm{k}) \\ \mathbf{\Delta}^*(- \bm{k}) & \mathbf{h}^*(- \bm{k})\end{array}\right)
    \end{equation}
    where $\mathbf{h}$ and $\mathbf{\Delta}$ are 2 by 2 matrix. Because of preservation of PHS, we only need to consider the positive frequency, which is what we can detect in experiment \cite{mag_topo_2_zero_freq}.

    Again, we have to diagonalize the $\bm{\eta} \mathbf{H}_{\mathrm{FF}}^{(')}$ by non-Hermitian Schrödinger-like equation:
    \begin{equation}
    i\hbar \frac{d}{dt} |\chi_{m}\rangle = \hbar \omega |\chi_{m}\rangle = \bm{\eta} \mathbf{H}_{\mathrm{FF}}^{(')} |\chi_{m}\rangle
    \end{equation}
    
     It is straightforward to verify that only the antiparallel
configuration violates CS
    \cite{Break_Chiral_1, Break_Chiral_2}   
    \begin{equation}\mathbf{\sigma_z} \mathbf{h} \mathbf{\sigma_z} \neq -\mathbf{h},\end{equation} which requires $\omega_l+D_l =0$ for any $k_x$.
    However, both cases break the TRS \cite{Break_TRS_1, Break_TRS_2} 
    \begin{equation}
    \bm{\eta}\hat{K}  \mathbf{H}_{\mathrm{FF}}^{(')}\hat{K}\bm{\eta} \neq  \mathbf{H}_{\mathrm{FF}}^{(')},
    \qquad
    \hat K:\text{ complex conjugation operator}.
    \end{equation}

    So we can use Chern number to classify the topological states. We refer to Ref.~\cite{Chern_3_topocourse, Chern_1_code}'s numerical method to calculate the Chern number. We note that, since $\bm{\eta} \mathbf{H}_{\mathrm{FF}}^{(')}$ is non-Hermitian, the inner product of two states should be defined as $\left \langle \ \cdot \ | \bm{\eta} | \ \cdot \ \right  \rangle$. We modify the methods based on Ref.~\cite{Chern_2_boson, Basic_Review} as follow:

    For a given right eigenstate 
\(
\bigl|\chi_{m}(\bm k)\bigr\rangle
\) 
of the non-Hermitian matrix $\bm{\eta} \mathbf{H}_{\mathrm{FF}}^{(')}$, the Berry connection of $m^{th}$ band must be defined with the $\bm{\eta}$–inner product:
    \begin{equation}
  \bm A_{m}(\bm k)=
  i\,\frac{%
      \bigl\langle\chi_{m}(\bm k)\bigr|
      \bm{\eta}\,
      \boldsymbol{\nabla}_{\bm k}
      \bigl|\chi_{m}(\bm k)\bigr\rangle}
     {%
      \bigl\langle\chi_{m}(\bm k)\bigr|
      \bm{\eta}\,
      \bigl|\chi_{m}(\bm k)\bigr\rangle}.
  \label{eq:eta_Berry_connection}
\end{equation}

Because the states are biorthonormalized such that  
\(
\langle\chi_{m}|\bm{\eta}|\chi_{m}\rangle=\pm1,
\)  
one immediately has  

\[
\langle\chi_{m}|\bm{\eta}|\chi_{m}\rangle=
\begin{cases}
 +1, & \text{magnon band},\\
 -1, & \text{antimagnon band}.
\end{cases}
\]

To keep track of the band, we order the
eigenvectors by the position of their largest‐magnitude component:

\begin{itemize}
  \item the {\it magnon-$a$} band in the first layer is identified by the eigenvector whose
        first component has the largest absolute value;
  \item the {\it antimagnon-$c$} band in the second layer is identified likewise via the
        fourth component, and so on.
\end{itemize}

Starting from the $\bm{\eta}$-metric Berry connection
$\bm A_{m}(\bm k)$ in
Eq.~(\ref{eq:eta_Berry_connection}),
the Berry curvature of the $m$-th band is

\begin{equation}
  \bm{\Omega}_{m}(\bm k)=
  \boldsymbol{\nabla}_{\bm k}
  \times
  \bm A_{m}(\bm k).
\end{equation}

Because the Bloch Hamiltonian depends only on
$(k_x,k_z)$, the sole non-vanishing component is

\begin{equation}
  \Omega^{y}_{m}(k_x,k_z)=
  i\!\left[
      \partial_{k_x}
        \langle\chi_{m}|\,
        \bm{\eta}\,
        \partial_{k_z}|\chi_{m}\rangle
      -
      \partial_{k_z}
        \langle\chi_{m}|\,
        \bm{\eta}\,
        \partial_{k_x}|\chi_{m}\rangle
    \right].
  \label{eq:Omega_y}
\end{equation}

The flux of this curvature over the two-dimensional BZ gives

\begin{equation}
  \Phi_{m}
  \;=\;
  \int_{\mathrm{BZ}}
  \mathrm d k_{x}\,
  \mathrm d k_{z}\,
  \Omega^{y}_{m}(k_x,k_z),
\end{equation}
and the Chern number is

\begin{equation}
  \mathrm{Ch}_{m}
  \;=\;
  \frac{\Phi_{m}}{2\pi}.
  \label{eq:Chern_def}
\end{equation}

If TRS were present,
the Berry curvature would satisfy
\(
  \Omega^{y}_{m}(-k_x,-k_z)=
 -\Omega^{y}_{m}(k_x,k_z)
\),
so that the integral in
Eq.~(\ref{eq:Chern_def}) vanishes and
\(\mathrm{Ch}_{m}=0\).
In our system, TRS is explicitly broken,
allowing for non-zero integer Chern numbers.

Directly differentiating numerically obtained eigenvectors
\(\bigl|\chi_{m}(\bm k)\bigr\rangle\)
is unreliable because each diagonalization adopts an arbitrary
phase (gauge).
Instead we follow the lattice algorithm of
Ref.~\cite{Chern_3_topocourse} and tile the BZ with
small squares
\(\Gamma_{n}\) (\(n=1,\dots,N\));  
the total Berry flux is then written as a sum of
plaquette fluxes:

\begin{equation}
  \Phi_{m}
  \;=\;
  \oint_{\mathrm{BZ}}
\mathrm d\bm k\!\cdot\!\bm A_{m}
  \;=\;
  \sum_{n}\,
  \oint_{\Gamma_{n}}
  \mathrm d\bm k\!\cdot\!\bm A_{m}
  \;\equiv\;
  \sum_{n}\phi_{n}.
  \label{eq:total_flux}
\end{equation}

Let the sum/product over the closed path of the plaquette $\Gamma_n$ be denoted by the primed sum/product
\(\sum_{\Gamma_{n}}^{\prime}\) or
\(\prod_{\Gamma_{n}}^{\prime}\)
(counter-clockwise order).
For a sufficiently fine mesh, one has

\begin{align}
  e^{i\phi_{n}}
  &=\exp\!\Bigl[i\oint_{\Gamma_{n}}
                 \mathrm d\bm k\!\cdot\!\bm A_{m}\Bigr]
      \approx
      \exp\!\Bigl[
        \sum_{\Gamma_{n}}^{\prime}
        \Delta\bm k\!\cdot\!\bm A_{m}
      \Bigr]                                 \notag\\[4pt]
  &\approx
      \prod_{\Gamma_{n}}^{\prime}
      \Bigl(1+\Delta\bm k\!\cdot\!\bm A_{m}\Bigr)
  \notag\\[4pt]
  &\approx
      \prod_{\Gamma_{n}}^{\prime}
      \frac{\langle\chi_{m}(\bm k)|
              \bm{\eta}|
              \chi_{m}(\bm k+\Delta\bm k)\rangle}
           {\langle\chi_{m}(\bm k)|
              \bm{\eta}|
              \chi_{m}(\bm k)\rangle}.
  \label{eq:plaquette_flux}
\end{align}

Here
\(\Delta\bm k\) is the momentum increment along the corresponding
edge and the overlaps are taken with the
$\bm{\eta}$-metric to maintain
pseudo-Hermiticity.  
Each factor in
Eq.~(\ref{eq:plaquette_flux})
is {gauge covariant}, while their product around a closed loop is
{gauge invariant}, eliminating any phase-singularity problem.

The Chern number finally becomes the sum of plaquette angles,

\begin{equation}
  \mathrm{Ch}_{m}
  \;=\;
  \frac{1}{2\pi}
  \sum_{n}
  \arg\!
    \left[
      e^{i\phi_{n}}
    \right]
  =
  \frac{1}{2\pi}
  \sum_{n}
  \arg
    \!\left[
      \prod_{\Gamma_{n}}^{\prime}
      \frac{\langle\chi_{m}(\bm k)|
              \bm{\eta}|
              \chi_{m}(\bm k+\Delta\bm k)\rangle}
           {\langle\chi_{m}(\bm k)|
              \bm{\eta}|
              \chi_{m}(\bm k)\rangle}
    \right].
  \label{eq:Chern_lattice}
\end{equation}

Equations
(\ref{eq:total_flux})–(\ref{eq:Chern_lattice})
constitute a fully gauge-invariant, numerically stable procedure
for evaluating the Chern numbers of the bosonic BdG bands.
    
    The simplified case had been discussed in Ref \cite{SSH1_Luqiao}, without considering $D_{li}$. In this manner, we can separate the magnon and antimagnon Hamiltonian and use the usual way to calculate the Chern number. However, the $D_{c2}$ perturbation breaks the CS of the 4 by 4 Hamiltonian. As a result, due to PHS, in the antiparallel case, if any band from \(a\) crosses any band from \(c\), the Chern number will be nonzero and represent a nontrivial state. Interestingly, the parallel configuration is also a 2D-SSH4 chain. If two bands are crossing, the Chern number is also non-zero. This effect cannot be predicted by the previous magnon-only picture. 
  
\section*{Section B: Topological Origin of MSSW} \label{sec:Appendix_B}
    \begin{figure}[t] 
    \capstart
    \centering 
    \includegraphics[width=0.5\textwidth]{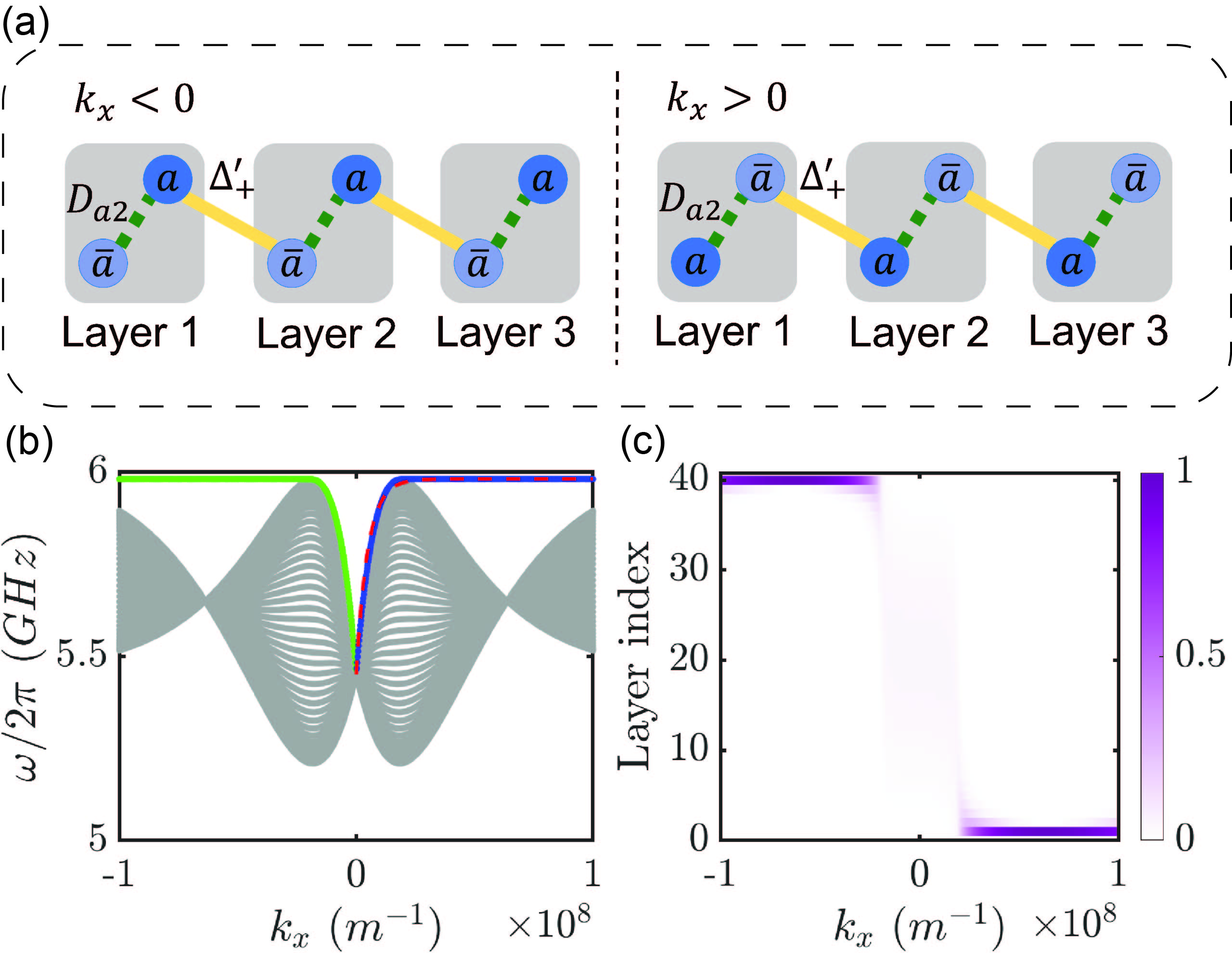} 
    \caption{\label{fig:10}(a)Illustration of the 2D-SSH chain for a bulk YIG insulator. (b) Band structure of MSSW in a $1\ \mathrm{\mu m}$-thick YIG film discretized into 40 layers, under an external magnetic field $B^y_a/\mu_0 = 1\times 10^5\ \mathrm{A/m}$. Bulk states are shown in gray. The green surface state (MSSW) propagates only along the upper surface, while the blue surface state (MSSW) propagates only along the lower surface. The red dashed line is the classical model prediction. (c) Spatial localization of these states, with darker purple indicating stronger localization.}
    \end{figure}
    In this section, we want to show that the famous MSSW  has the same topological origin as our model. We first show the 2D-SSH model for a thin FM insulator and explain the mechanism. Then we show the calculated band structure of MSSW. 
    
    MSSW was originally described phenomenologically by coupling the Landau–Lifshitz–Gilbert (LLG) equation with Maxwell’s equations\cite{Magnetization_1st_ed}. Recent studies, however, have revealed that dipolar surface spin waves in FM possess a topological origin when TRS is preserved \cite{Topo_MSSW_2019}. Here, we extend this framework to FM insulators with broken TRS by an external magnetic field, where the Chern number serves as the topological invariant. The model is similar to the FM parallel configuration, but with a single type of insulator, a larger number of layers (\(M = 40\) in our calculation), and vanishing exchange interaction. The nearest-neighbor approximation is applied. The two-dimensional SSH model used is illustrated in Fig.\ref{fig:10}(a).

    To elucidate the directional propagation of MSSWs on the upper and lower surfaces, we note that for $k_x<0$ the magnon surface state is localized on the uppermost layer, while for $k_x>0$ it is localized on the lowermost layer. This explains the nonreciprocal behavior of MSSWs and demonstrates that the topological surface states in both our FM multilayer and AFM/FM models share the same physical origin as the MSSW. In Fig.\ref{fig:10}(b), we present the calculated band structure of a $1\mathrm{\mu m}$ YIG thin film discretized into 40 layers. Bulk states are highlighted in green and blue, while the spatial localization of the states is shown in Fig.\ref{fig:10}(c). The surface-localized modes exhibit the characteristic asymptotic frequency predicted for MSSW \cite{MSSW_First}. The result is consistent with the phenomenological prediction $\omega^2 = \left( \omega_{\mathrm{H}} + \frac{\omega_{\mathrm{M}}}{2} \right)^2 - \left( \frac{\omega_{\mathrm{M}}}{2} \right)^2 \exp(-2k_xd_{\mathrm{ph}})$ \cite{Magnetization_1st_ed}, which shown in red dash line in Fig. \ref{fig:10}(b). Here $\omega_{\mathrm{H}} = \gamma\mu_0 H^y_a =\gamma B^y_a$ and $\omega_{\mathrm{M}} = \gamma\mu_0M_a$. These results provide a clear topological interpretation of the MSSW in FM insulators.  
    
    \added[id=YF]{However, truncating the dipolar interaction to nearest neighbors limits the model’s accuracy in the thickness. This approximation underestimates the effective thickness of the film.  To compensate, we introduce a phenomenological thickness  $d_{\mathrm{ph}} =0.1\mathrm{\mu m}$, which corresponds to four times the single-layer thickness ($d$) once the stack contains many layers, i.e., $d_{ph} \approx4d$. This empirical scaling effectively captures the long-range nature of dipolar interactions in multilayer systems.
    With this adjustment, a 40-layer discretization of a $1\mathrm{\mu m}$-thick YIG film yields a spin-wave spectrum nearly identical to those obtained using 50 or 60 layers. This "four-layer rule" restores the missing dipolar weight with minimal computational overhead, enabling accurate modeling of MSSWs in thick ferromagnetic films.}
\section*{Section C: Calculation of Band Structure of AFM/FM Multilayer} \label{sec:Appendix_C}
    In this section, we follow a similar procedure in Sec. A to calculate the magnonic band structure in AFM/FM multilayer. We start from the AFM/FM bilayer Hamiltonian and then extend it to the multilayer configuration. Finally, we arrive at the multilayer Hamiltonian.
    
    Consider we have an AFM/FM bilayer. We follow a similar procedure. The total Hamiltonian can be expressed as:
    \begin{equation}
      \hat{\mathcal H}_{\mathrm{AF}}
      =\hat{\mathcal H}_{a\mathrm{0}}+\hat{\mathcal H}_{bc\mathrm{0}}+\sum_{l=a,b,c}\hat{\mathcal H}_{l\mathrm{D0}}
      +\sum_{l\neq l'}\hat{\mathcal H}_{ll'\mathrm{D}},
    \end{equation}
    with 
    \begin{equation}
    \begin{aligned} &
    \hat{\mathcal H}_{a\mathrm{0}} = -g\mu_{\mathrm B}({S}_{a} {B}_{a}^{y}N_{a}-\sum_{\bm k}({B}_{a}^{y}\hat{a}_{\bm k}^{\dagger}\hat{a}_{\bm k}))
     +{S}_{a}^{2} {J}_{a}N_{a}-\sum_{\bm{k}}2S_{a}J_{a}\hat{a}_{\bm{k}}^{\dagger}\hat{a}_{\bm{k}}\sum_{i=x,y,z}\sin^{2}(\frac{a_0k_{i}}{2})
     \end{aligned}
    \end{equation}
\begin{align}
\hat{\mathcal H}_{bc0}
&=
-g\mu_{\mathrm B}\sum_{l=b,c}\sum_{i}
  \hat{\bm S}_{li}\!\cdot\!\bm B_{l}
\;+\;
J_{bc}
\sum_{\langle i;j\neq i\rangle}
  \hat{\bm S}_{bi}\!\cdot\!\hat{\bm S}_{cj}.
\end{align}
\begin{align}
\hat{\mathcal H}_{l\mathrm D0}
&=
\frac{\mu_{0}(g\mu_{\mathrm B})^{2}}{2}
\sum_{i\neq j}
\frac{%
       R_{ij}^{2}\,
      \hat{\bm S}_{li}\!\cdot\!\hat{\bm S}_{lj}
      -3(\bm R_{ij}\!\cdot\!\hat{\bm S}_{li})
        (\bm R_{ij}\!\cdot\!\hat{\bm S}_{lj})
     }{%
      R_{ij}^{5}
     }.
\end{align}
\begin{align}
\hat{\mathcal H}_{ll'\mathrm D}
&=
\frac{\mu_{0}(g\mu_{\mathrm B})^{2}}{2}
\sum_{\substack{i\in l\, j\in l'}}
\frac{R_{ij}^{2}\,
      \hat{\bm S}_{li}\!\cdot\!\hat{\bm S}_{l'j}
      -3(\bm R_{ij}\!\cdot\!\hat{\bm S}_{li})
        (\bm R_{ij}\!\cdot\!\hat{\bm S}_{l'j})
     }{%
       R_{ij}^{5}
     }.
\end{align} 
    Atom \(a\) is from FM layer, and \(b\) and \(c\) type atoms are from the AFM layer. We assume the spins of atom $a$ and $b$ are in the same direction,i.e., the $+y$ direction. Spin of atom \(c\) is in the $-y$ direction. We adopted the results from the previous section for $\hat{\mathcal H}_{a\mathrm{0}}$, which is the Hamiltonian of Zeeman interaction and intralayer exchange interaction of layer $a$. Similarly, $\hat{\mathcal H}_{bc0}$ is the Hamiltonian of Zeeman interaction and intralayer exchange interaction of AFM layer, which has \(b\) and \(c\) type atoms. In this case, we ignore the FM exchange interaction in AFM layer of \(b\) and \(c\) type atoms to the same type and only consider the AFM exchange interaction between \(b\) and \(c\), i.e. $J_{bc}>0$ for AFM exchange coupling. $\hat{\mathcal H}_{l\mathrm D0}$ is the Hamiltonian of dipolar interaction of the same type of atom, therefore only representing the intralayer dipolar interaction. $\hat{\mathcal H}_{ll'\mathrm D}$ is the Hamiltonian of dipolar interlayer interaction of different types of atoms. $\hat{\mathcal H}_{ab\mathrm D}$ and $\hat{\mathcal H}_{ac\mathrm D}$ represent both interlayer dipolar interaction while $\hat{\mathcal H}_{bc\mathrm D}$ still represents intralyer dipolar interaction.

    We first focus on $\hat{\mathcal H}_{bc0}$. The HP transformation for \(a\) and \(c\) remains the same, and for \(b\) is the same as \(a\). Through a similar procedure, we arrive at
\begin{align}
\hat{\mathcal H}_{bc0}&=g\mu_{\mathrm B}\Bigl[N_{b}B_{0}^{y}(-S_{b}+S_{c})+B_{0}^{y}\sum_{\bm{k}}(\hat{b}_{\bm{k}}^{\dagger}\hat{b}_{\bm{k}}-\hat{c}_{\bm{k}}^{\dagger}\hat{c}_{\bm{k}})\Bigr]-8J_{bc}N_{b}S_{b}S_{c}    
\notag\\[4pt]
&\quad
+J_{bc}\sum_{\bm{k}}\Bigl\{8S_{b}\hat{c}_{\bm{k}}^{\dagger}\hat{c}_{\bm{k}}+8S_{c}\hat{b}_{\bm{k}}^{\dagger}\hat{b}_{\bm{k}}+\sqrt{S_{b}S_{c}}\bigl[\hat{b}_{\bm{k}}\hat{c}_{-\bm{k}}\sum_{\bm{\delta}}e^{i\bm{k}\!\cdot\!\bm{\delta}}+\hat{b}_{\bm{k}}^{\dagger}\hat{c}_{-\bm{k}}^{\dagger}\sum_{\bm{\delta}}e^{-i\bm{k}\!\cdot\!\bm{\delta}}\bigr]\Bigr\}.
\end{align}
    Here $\bm{\delta}$ is the nearest neighbor displacement between atom \(b\) and \(c\). 
    
    Next, we focus on the dipolar interaction for the same type of atoms. We can adopt the results derived for the case of FM layers:

\begin{align}
\sum_{l=a,c}\hat{\mathcal H}_{l\mathrm D0}
&=
\frac{\pi\mu_{0}(g\mu_{\mathrm B})^{2}}{2}\,
\sum_{l=a,c} d_{l}S_{l}
\int\!\frac{\mathrm{d}k_{x}}{2\pi} 
\Bigl[
      \bigl(
         \hat{l}_{k_{x}}^{\dagger}\hat{l}_{k_{x}}
        +\hat{l}_{k_{x}}\hat{l}_{k_{x}}^{\dagger}
      \bigr)
      +\bigl(
         \hat{l}_{k_{x}}^{\dagger}\hat{l}_{k_{x}}^{\dagger}
        +\hat{l}_{k_{x}}\hat{l}_{k_{x}}
      \bigr)\,(2\Phi_{l}-1)
\Bigr]
+ \mathrm{H.c.}
\end{align}
    Finally, we will derive the dipolar interaction between different types of atoms. For dipolar interaction between atom \(a\) and \(b\) or \(a\) and \(c\), we can still adopt the results from above because both  \(b\) and \(c\)'s lattice type is simple cubic \cite{AFM_Review}:
\begin{align}
\hat{\mathcal H}_{ac\mathrm D}
=
\frac{\mu_{0}(g\mu_{\mathrm B})^{2}\pi}{2}\,
\sqrt{S_{a}S_{c}}\,
\int\!\frac{\mathrm d k_{x}}{2\pi}          
 \Phi_{a}\Phi_{c}
\Bigl[
   (\lvert k_{x}\rvert-k_{x})\,
   \hat{c}_{k_{x}}^{\dagger}\hat{a}_{k_{x}}
   +(\lvert k_{x}\rvert+k_{x})\,
   \hat{c}_{k_{x}}\hat{a}_{k_{x}}^{\dagger}
\Bigr]
+\mathrm{H.c.}
\end{align}

\begin{align}
\hat{\mathcal H}_{ab\mathrm D}
=
\frac{\mu_{0}(g\mu_{\mathrm B})^{2}\pi}{2}\,
\sqrt{S_{a}S_{b}}\,
\int\!\frac{\mathrm d k_{x}}{2\pi}  \Phi_{a}\Phi_{b}  
\Bigl[
   (-\lvert k_{x}\rvert-k_{x})\,
   \hat{b}_{k_{x}}^{\dagger}\hat{a}_{k_{x}}
   +(-\lvert k_{x}\rvert+k_{x})\,
   \hat{b}_{k_{x}}\hat{a}_{k_{x}}^{\dagger}
\Bigr]
+\mathrm{H.c.}
\end{align}
    For $\hat{\mathcal H}_{bc\mathrm D}$, we simply replace one of $\bm{S}_l$ to $\bm{S}_{l'}$ in the expression of $ \hat{\mathcal H}_{l\mathrm D0}$. This is valid because we assume our lattice for AFM is also a simple cubic but with a basis of 2 atoms, with a $b$ atom in the center of the conventional cell. So, we arrive at: 
\begin{equation}
\hat{\mathcal H}_{bc\mathrm D}
=
\frac{\mu_{0}(g\mu_{\mathrm B})^{2}}{4}
\int\!\frac{\mathrm d k_{x}\,\mathrm d k_{y}}{(2\pi)^{2}}\;
\hat{\bm S}_{b}^{\mathsf T}(\bm{k}_{\parallel})\,
 \mathbf G_{bc}(\bm{k}_{\parallel})\,
\hat{\bm S}_{c}(\bm{k}_{\parallel})
+\mathrm{H.c.}
\end{equation}

    where 
    \begin{equation}
 \mathbf G_{bc}(\bm{k}_{\parallel})
=
\frac{4\pi d_{l}}{k_{\parallel}^{2}}
\begin{bmatrix}
k_{x}^{2}(1-\Phi_{l}) & k_{x}k_{y}(1-\Phi_{l}) & 0\\
k_{x}k_{y}(1-\Phi_{l}) & k_{y}^{2}(1-\Phi_{l}) & 0\\
0                      & 0                      & k_{\parallel}^{2}\Phi_{l}
\end{bmatrix}.
\end{equation}
    again assume $k_y =0$, we get
     \begin{equation}
\hat{\mathcal H}_{bc\mathrm D}
=\frac{\pi\mu_{0}(g\mu_{\mathrm B})^{2}}{2}\,
  d_{c}\sqrt{S_{b}S_{c}}
  \int\!\frac{\mathrm d k_{x}}{2\pi}
  \Bigl[
      (\hat{b}_{k_{x}}^{\dagger}\hat{c}_{k_{x}}
       +\hat{b}_{k_{x}}\hat{c}_{k_{x}}^{\dagger})
      (2\Phi_{l}-1)
      -\bigl(
          \hat{b}_{k_{x}}^{\dagger}\hat{c}_{k_{x}}^{\dagger}
         +\hat{b}_{k_{x}}\hat{c}_{k_{x}}
       \bigr)
  \Bigr]
  +\mathrm{H.c.}
\end{equation}
    Again, by defining 
    \begin{equation}
\label{eq:Nambu_basis_AF}
\hat{\bm{\beta}}_{k_{x}}\equiv
\bigl[\,
  \hat a_{k_{x}}\;,\,
  \hat b_{k_{x}}\;,\,
  \hat c_{k_{x}}\;,\,
  \hat a_{-k_{x}}^{\dagger}\;,\,
  \hat b_{-k_{x}}^{\dagger}\;,\,
  \hat c_{-k_{x}}^{\dagger}
\bigr]^{\mathsf T},
\end{equation}
We can write the total Hamiltonian in matrix form:
    \begin{equation}
\hat{\mathcal H}_{\mathrm{AF}}
=\frac{1}{2}\sum_{k_{x}}
\hat{\bm{\beta}}_{k_{x}}^{\dagger}\,
 \mathbf{H}_{\mathrm{BdG}}^{\mathrm{AF}}\,
\hat{\bm{\beta}}_{k_{x}}.
\end{equation}
    where
    \begin{equation}
    \mathbf{H}_{\mathrm{BdG}}^{\mathrm{AF}}=\begin{bmatrix}
    \omega_a+D_{a1}&  0&  \Delta_{ac+}&  D_{a2}&  \Delta_{ab+}&  0\\
    0&  \omega_b+D_{b1}&  \Delta_{bc1}&  \Delta_{ab-}&  D_{b2}& A_{c}+\Delta_{bc2}\\
    \Delta_{ac+}&  \Delta_{bc1}&  \omega_c+D_{c1}&   0&  A_{b}+\Delta_{bc2}& D_{c2}\\
    D_{a2}&  \Delta_{ab-}&   0&  \omega_a+D_{a1}&  0& \Delta_{ac-}\\
    \Delta_{ab+}&  D_{b2}&  A_{c}+\Delta_{bc2}&  0&  \omega_b+D_{b1}& \Delta_{bc1}\\
     0&  A_{b}+\Delta_{bc2}&  D_{c2}&  \Delta_{ac-}&  \Delta_{bc1}&  \omega_c+D_{c1}
    \end{bmatrix},
    \end{equation}
    with
    \begin{equation}
    \omega_a = S_{a}J_{a}a_0^2k_x^2 \ + \ g\mu_{\mathrm B}B_{a}^y
    \end{equation}
    \begin{equation}
    \omega_b = g\mu_{\mathrm B}B_{b}^y \ +\ 8J_{bc}S_c
    \end{equation}
    \begin{equation}
    \omega_c = -g\mu_{\mathrm B}B_{c}^y \ +\ 8J_{bc}S_b
    \end{equation}
    \begin{equation}
    A_l = 8J_{bc}S_lcos(\frac{k_xa_0}{2})
    \end{equation}  
    \begin{equation}
    D_{l1} = \pi\mu_0(g\mu_{\mathrm B})^2d_lS_l
    \end{equation}
    \begin{equation}
    D_{l2} = \pi\mu_0(g\mu_{\mathrm B})^2d_lS_l(-2\Phi_l+1)
    \end{equation}
    \begin{equation}
    \Delta_{ac\pm } = \pi\mu_0(g\mu_{\mathrm B})^2\sqrt{S_aS_c}\Phi_a\Phi_cd_ad_c(|k_x|\pm k_x)
    \end{equation}
    \begin{equation}
    \Delta_{ab\pm } = \pi\mu_0(g\mu_{\mathrm B})^2\sqrt{S_aS_b}\Phi_a\Phi_bd_ad_c(-|k_x|\pm k_x)
    \end{equation}
    \begin{equation}
    \Delta_{bc1} = \pi\mu_0(g\mu_{\mathrm B})^2 d_c \sqrt{S_bS_c} (2\Phi_l-1)
    \end{equation}
    \begin{equation}
    \Delta_{bc2} = -\pi\mu_0(g\mu_{\mathrm B})^2 d_c \sqrt{S_bS_c}
    \end{equation}

    Again, for the bosonic system, we solve for the eigenvalues of
    $\bm{\eta}\mathbf{H}_{\mathrm{BdG}}^{\mathrm{AF}}$, where
    \[
    \bm{\eta}\equiv\mathrm{diag}\,(1,1,1,-1,-1-1).
    \] to obtain the bilayer band structure. To obtain more layers' band structure, we just adopt OBC and follow the above formalism to diagonalize the larger BdG Hamiltonian.
    
    Similarly, we construct a two-dimensional SSH-like model by stacking the AFM/FM bilayer unit to a multilayer.  The resulting structure is equivalent to two 2D-SSH4 chains—one with parallel and one with antiparallel FM orientations—hybridized via the AFM exchange coupling $J_{bc}$.  Like the previous case, there exist nonreciprocal surface modes on the upmost and bottommost layers propagating in a single direction only.

    Assume in each layer $k_z=0$, meaning in each layer, spins in the $z$ direction are uniform and not propagating. If there are $M$ bilayers in total, define the magnon state of each layer to be $\left | n, l  \right \rangle$ where $n=1,...,M$ and $l = a,b,c$. Hence, the periodic boundary condition along the $z$ direction reads: 
    \begin{equation}
    \left | M+1, l  \right \rangle = \left | 1, l  \right \rangle 
    \end{equation}
    And the Fourier Transform for the magnon state along the $z$ direction gives: 
    \begin{equation}
    \left | n, l  \right \rangle = \frac{1}{\sqrt{M}}\sum_{k_z}^{}e^{-in k_z (d_a+d_c)}\left | k_z, l  \right \rangle 
    \end{equation}
    $k_z$ here means the wave propagating between layers instead of within layers.

    Again, define $d=d_a+d_c$. After adopting the periodic condition along the $z$ direction, the 6 by 6 Hamiltonian $\mathbf{H}_{\mathrm{AF}}$ can be constructed by combining $\mathbf{H}$ and $\mathbf{H}'$ in the previous section. 
    \begin{equation}
    \hspace{-1cm}
    \mathbf{H}_{2}=\begin{bmatrix}
    \omega_a+D_{a1}&  0&  \Delta_{ac+} +e^{-ik_zd}\Delta_{ac-} &  D_{a2}&  \Delta_{ab+}  +e^{-ik_zd}\Delta_{ab-}  & 0\\
    0&  \omega_b+D_{b1}&  \Delta_{bc1}&  \Delta_{ab-}  +e^{ik_zd}\Delta_{ab+}  &  D_{b2}& A_{c}+\Delta_{bc2}\\
    \Delta_{ac+} +e^{ik_zd}\Delta_{ac-} &  \Delta_{bc1}&  \omega_c+D_{c1}&  0&  A_{b}+\Delta_{bc2}& D_{c2}\\
    D_{a2}&  \Delta_{ab-} +e^{-ik_zd}\Delta_{ab+}  &  0&  \omega_a+D_{a1}&  0& \Delta_{ac-} +e^{ik_zd}\Delta_{ac+} \\
    \Delta_{ab+} +e^{ik_zd}\Delta_{ab-}  &  D_{b2}&  A_{c}+\Delta_{bc2}&  0&  \omega_b+D_{b1}& \Delta_{bc1}\\
    0&  A_{b}+\Delta_{bc2}&  D_{c2}&  \Delta_{ac-} +e^{-ik_zd}\Delta_{ac+} &  \Delta_{bc1}&  \omega_c+D_{c1}
    \vspace{4pt}
    \end{bmatrix}.
    \vspace{4pt}
    \end{equation}
    
     Again, we assume: $gu_B/\hbar = 1.76 \times 10^{11} \ \mathrm{rad/T}$,
     $S_a\hbar = 7.95 \times 10^{-7} \ \mathrm{J\,s}$, 
     $S_c\hbar = 4.20 \times 10^{-6} \ \mathrm{J\,s}$, 
     $d_a=d_c=10\ \mathrm{nm}$, 
     $-S_{a}J_{a}a_0^2 = 9.30\times 10^{-6}\ \mathrm{m^2/s}$ and 
     $J_{bc}S_c= J_{bc}S_b =2.76\times 10^{10} \ \mathrm{1/s}$.  
     AFM/FM multilayer structures are well-suited for experimental realization, as the exchange coupling in AFM layers typically exceeds that of FM layers. This results in intrinsic band crossings in few $\mathrm{GHz}$ range, enabling the observation of nontrivial topological states in microwave systems.
     
    We obtain the band structure by diagonalize $\bm{\eta}\mathbf{H}_{\mathrm{AF}}$. It satisfies the PHS while breaking the TRS and CS. We can calculate the Chern number for each band by the formalism in Section A.

\section*{Section D: Simulation of Band Structure of FM Multilayer and Chirality} \label{sec:Appendix_D}
    \begin{figure}[b] 
    \capstart
    \centering 
    \includegraphics[width=0.65\textwidth]{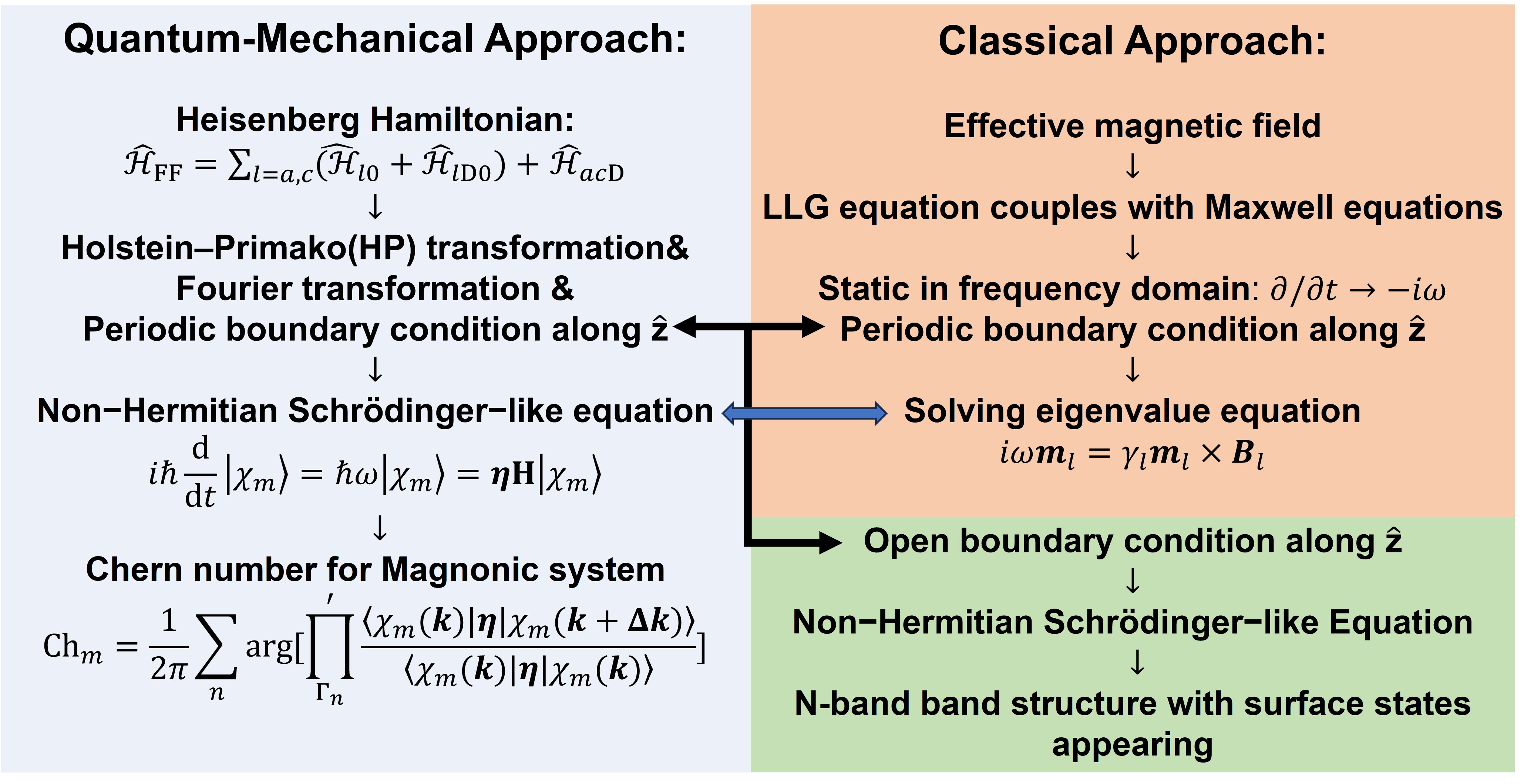} 
    \caption{\label{fig:12}Summary of the quantum and classical formalism.}
    \end{figure}

    \begin{figure}[t] 
    \capstart
    \centering 
    \includegraphics[width=\textwidth]{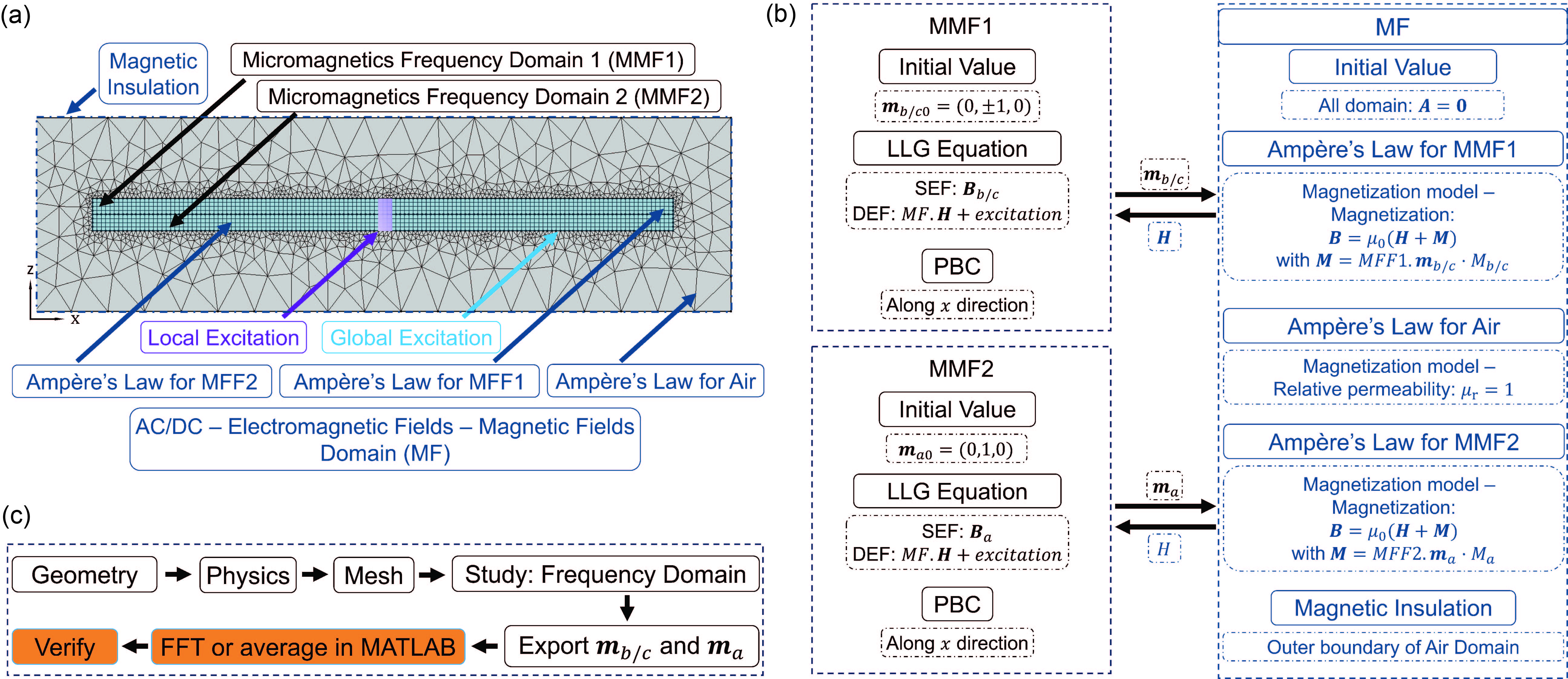} 
    \caption{\label{fig:11}(a) The two-dimensional model with corresponding mesh layout used for numerical simulations. (b) Schematic diagram illustrating the multiphysics coupling approach employed to simulate dipolar interactions between magnetic layers. (c) Flow chart of the simulation procedure; steps highlighted in orange are performed using MATLAB.}
    \end{figure}
    
    \begin{figure}[t] 
    \capstart
    \centering 
    \includegraphics[width=0.5\textwidth]{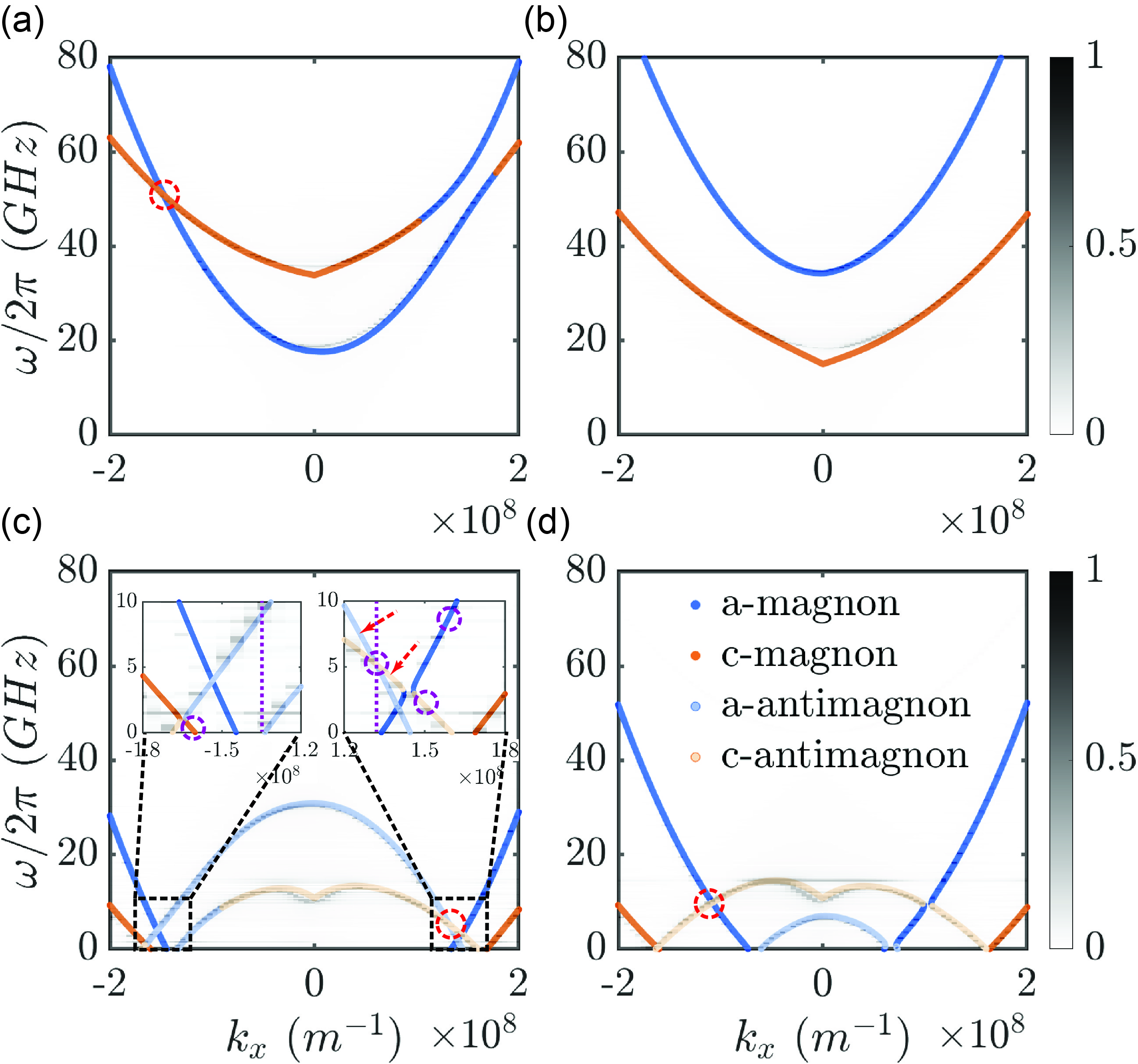} 
    \caption{\label{fig:6}(a)-(d) Band structures of one antiparallel FM bilayer unit. The simulation results are presented as color maps, with the color bar representing the lowest to highest spin-wave modes within each individual simulation. Topological surface states are in red circles and red arrows. The overall effective external fields of each layer are (a) $B_a^y/\mu_0 = 4.78\times 10^5 \mathrm{A/m}$ and $B_c^y/\mu_0 = -6.85\times 10^5 \mathrm{A/m}$ (nontrivial) (b) $B_a^y/\mu_0 = 9.28\times 10^5 \mathrm{A/m}$ and $B_c^y/\mu_0 = -2.35\times 10^5 \mathrm{A/m}$ (trivial) (c) $B_a^y/\mu_0 = -9.5\times 10^5 \mathrm{A/m}$ and $B_c^y/\mu_0 = 8.5\times 10^5 \mathrm{A/m}$ (nontrivial) (d) $B_a^y/\mu_0 = -2.8\times 10^5 \mathrm{A/m}$ and $B_c^y/\mu_0 = 8.5\times 10^5 \mathrm{A/m}$ (nontrivial). The subplot in (c) marked additional simulation points corresponding to Fig. \ref{fig:7} by purple dashed lines and Fig. \ref{fig:8} -\ref{fig:9} by purple circles.}
    \end{figure}

    \begin{figure}[t] 
    \capstart
    \centering 
    \includegraphics[width=0.5\textwidth]{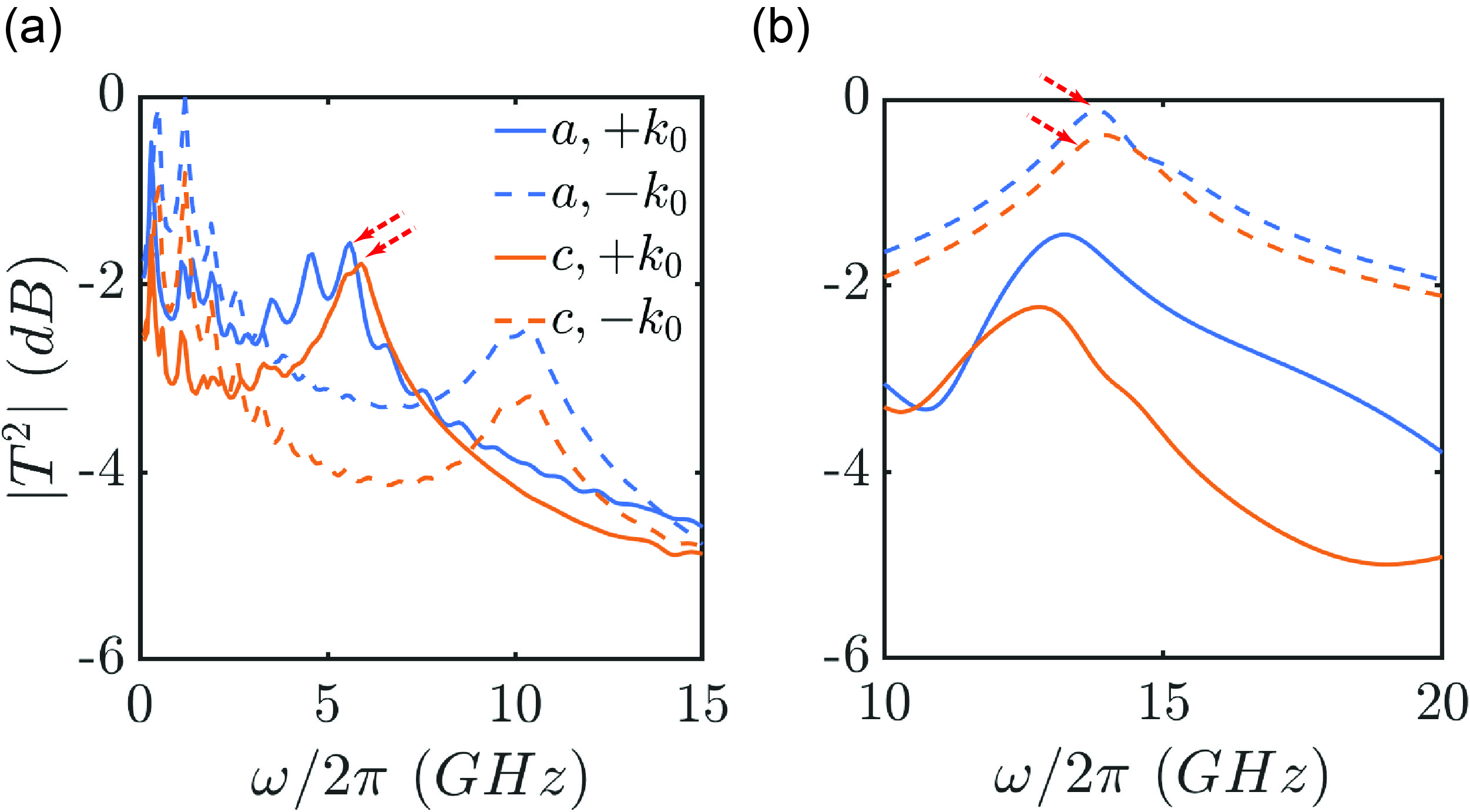} 
    \caption{\label{fig:7}(a)-(b) Simulation on propagating spin-wave spectroscopy of transmission rate of linear excitation. The peaks indicate the presence of magnon or antimagnon states at specific frequencies. The red arrows indicate the nonreciprocal topological surface states. (a) $k_0=1.32 \times 10^8 m^{-1}$, indicating in subplot Fig. \ref{fig:6}(c) by purple dash line. (b) $+k_0=0.616 \times 10^8 m^{-1}$, $-k_0=-0.72 \times 10^8 m^{-1}$ indicating in subplot Fig. \ref{fig:2}(c) by purple dash line.}
    \end{figure}

    \begin{figure*}[t] 
    \capstart
    \centering 
    \includegraphics[width=\textwidth]{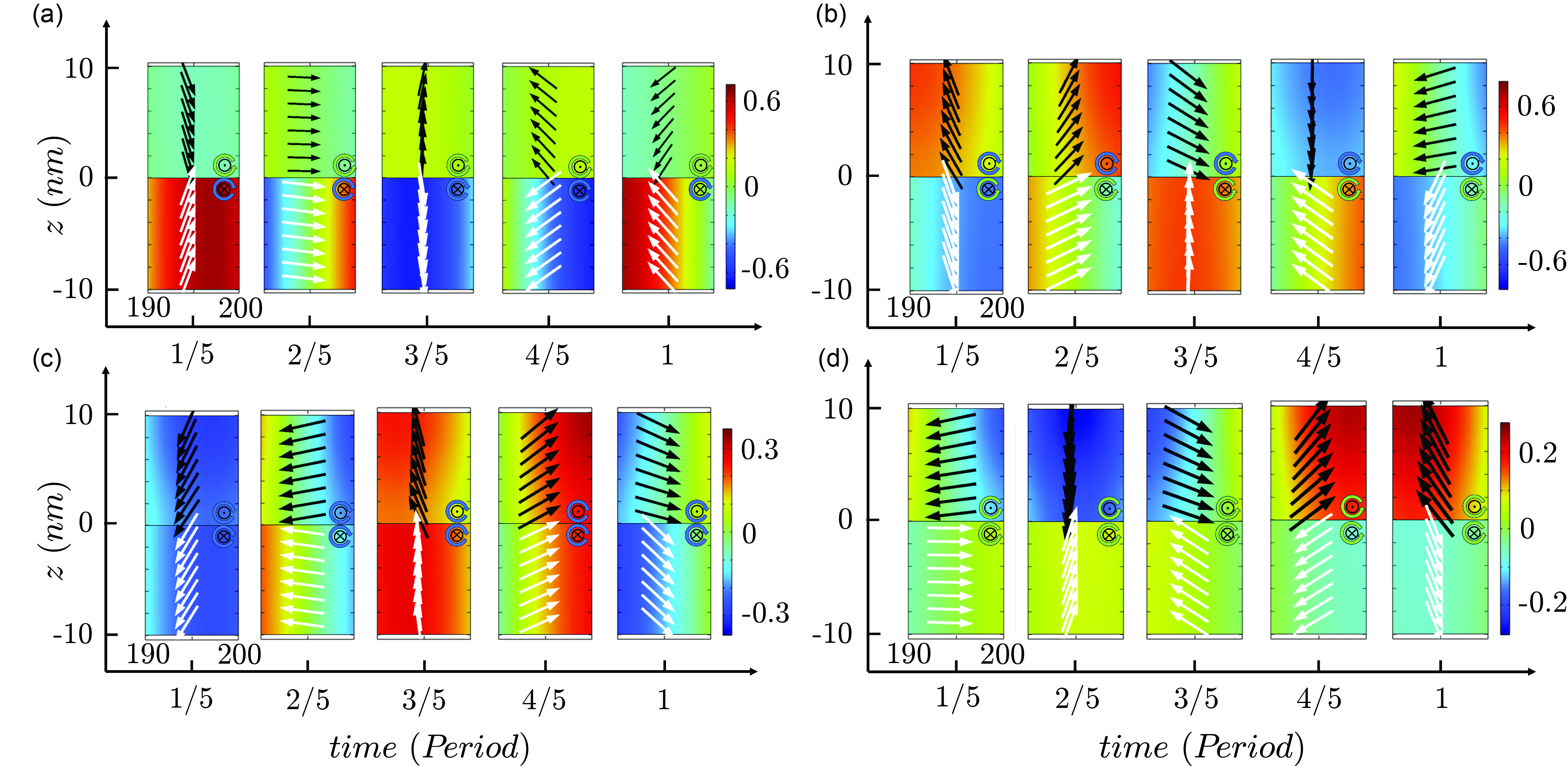} 
    \caption{\label{fig:8}(a)-(d) Simulation of chirality configurations in the time domain at times equal to $T/5$, $2T/5$, $3T/5$, $4T/5$, $T$, where $T$ is the period of the spin wave. Black and white arrows depict the magnetic moment directions of the $FM_a$ and $FM_c$ layers projected onto the $x$-$z$ plane, with their lengths scaled for clarity. The color map represents the magnitude of $z$ component of unit magnetic moments in each layer. The chirality of the spin wave in each layer under excitation is drawn by green circular arrows representing RH and blue representing LH. The black arrows along the $y$ direction represent the direction of the magnetization in each layer. The plots are all in the $x-z$ plane, with the section from $190\ nm$ to $200\ nm$ in $x$ direction and from $-10\ nm$ to $10\ nm$ in $z$ direction . The excitation is linearly polarized. The corresponding states are marked in subplots in Fig. \ref{fig:6}(c) by purple circles. (a) $k_1=1.59\times10^8  m^{-1}$. (b) $k_1=1.32\times10^8  m^{-1}$ (topological). (c) $k_1=1.49\times10^8  m^{-1}$. (d) $ k_1=-1.62\times10^8  m^{-1}$.}
    \end{figure*}

    \begin{figure}[t] 
    \capstart
    \centering 
    \includegraphics[width=0.5\textwidth]{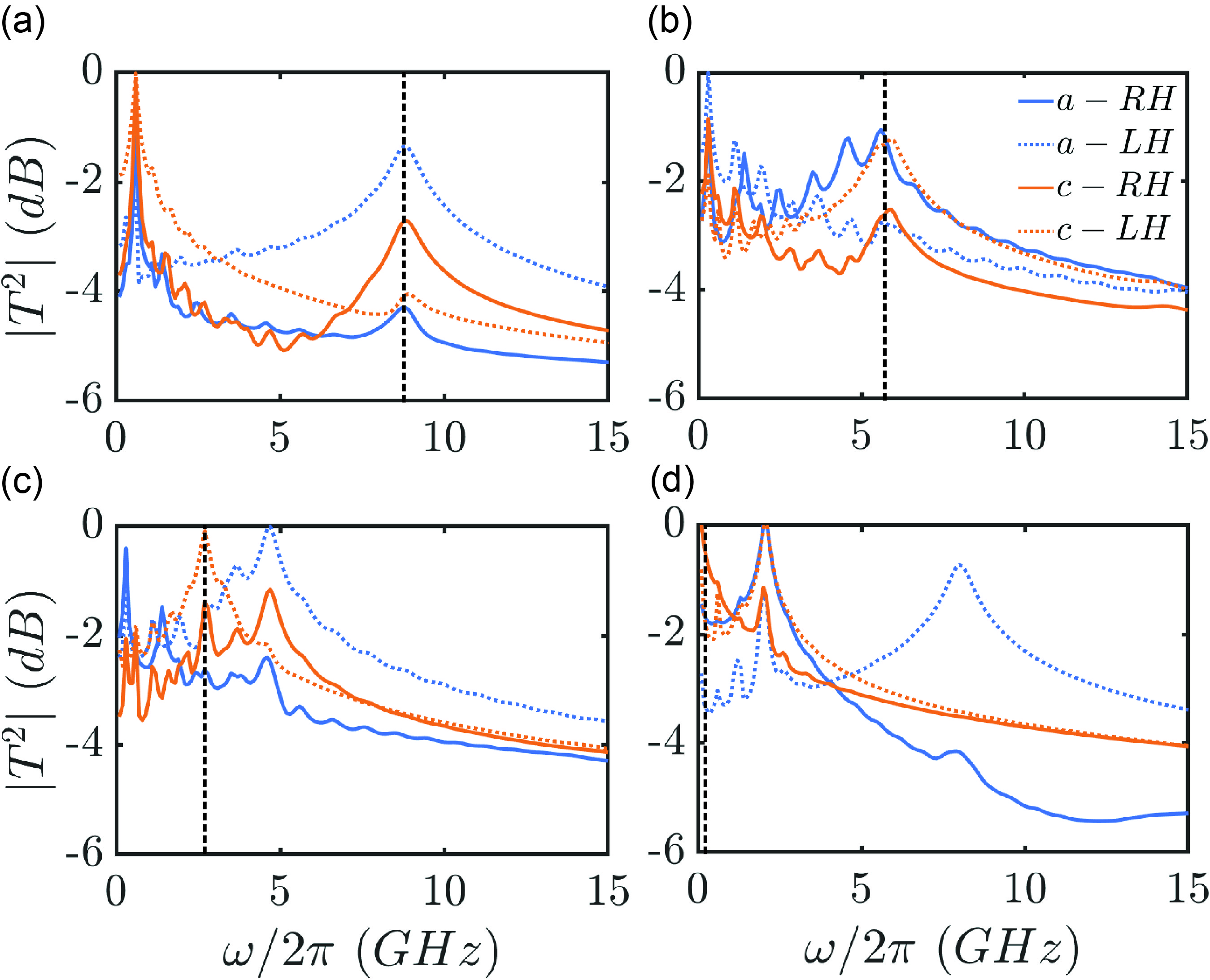} 
    \caption{\label{fig:9}(a)-(d) Simulation of circular mode response of linear excitation at a fixed wave vector $k_1$ of propagating spin wave in $x$ direction. The corresponding states are marked in subplots in Fig. \ref{fig:6}(c) by purple circles. The black dashed lines indicate the desired states. (a) $k_1=1.59\times10^8 \ m^{-1}$ and $\omega/2\pi = 8.8\ \mathrm{GHz}$. (b) $k_1=1.32\times10^8 \ m^{-1}$ and $\omega/2\pi = 5.8\ \mathrm{GHz}$. (c) $k_1=1.49\times10^8 \ m^{-1}$ and $\omega/2\pi = 3.3\ \mathrm{GHz}$ (topological). (d) $ k_1=-1.62\times10^8 \ m^{-1}$ and $\omega/2\pi = 0.4\ \mathrm{GHz}$.}
    \end{figure} 

    In this section, we first show the consistency between the quantum mechanical method and the classical method based on the LLG equation. We then demonstrate the classical method for simulating magnonic band structure and propagating spin-wave spectroscopy. Finally, we simulate the chirality of the bilayer FM unit cell and show that all four possible chirality configurations are achieved at a few $\mathrm{GHz}$, with one topologically protected.
    
    To verify the calculation, we do the simulation in COMSOL Multiphysics by Micromagnetic Module \cite{COMSOL1_MM, COMSOL2_acdc}. However, the simulation is based on the LLG equation and Maxwell equations, which are all classical. Hence, we have to map the second quantization language back to classical language \cite{Basic_Review}. We summarize the quantum-mechanical approach and classical approach we used in Fig. \ref{fig:12}. We begin by introducing the classical approach. The time domain LLG equation is given by \cite{SOT3_formalism}:

    \begin{equation}
    \frac{\partial \bm{m}_{l}}{\partial t}=-\gamma_l \bm{m}_{l} \times  \bm{B}+\alpha_{l} \bm{m}_{l} \times \frac{\partial \bm{m}_{l}}{\partial t}+\bm{\tau}_{\mathrm{D L}, l}+\bm{\tau}_{\mathrm{F L},l} \,,
    \end{equation}
    \begin{equation}
    \bm{\tau}_{\mathrm{D L},l}\equiv \gamma_{l}  B_{\mathrm{D L}} \bm{m}_{l} \times \bm{\sigma} \times \bm{m}_{l}, 
    \end{equation}
    \begin{equation}
    \bm{\tau}_{\mathrm{F L},l}\equiv \gamma_{l}  B_{\mathrm{F L}} \bm{m}_{l} \times \bm{\sigma},
    \end{equation}
    \begin{equation}
    B_{\mathrm{D L}} = \frac{J_{e}\theta_{\mathrm{SH}}\hbar}{2|e|d_l M_{l}} ,
    \end{equation}    
    where $\bm{B}$ is the effective magnetic field, and $\bm{m}_l$ is the magnetization unit vector in layer $l$, defined as $\bm{m}_l = \bm{M}_l / M_l$. $M_l$ is the saturation magnetization of layer $l$. $\alpha_l$ is the Gilbert damping constant in layer $l$, and $\bm{\sigma}$ is the spin-polarization direction of spin-orbit torque, which is in $y$ or $-y$ direction indicated by the purple arrows in Fig. \ref{fig:1}. $|e|$ is the charge unit, and $\theta_{SH}$ is spin Hall angle. We also have $B_{\mathrm{F L}}=\eta B_{\mathrm{D L}}$ where $\eta$ describes the ratio of these two fields. $\bm{\tau}_{\mathrm{D L},l}$ and $\bm{\tau}_{\mathrm{F L},l}$ are damping-like and field-like torques of the layer $l$. We define the overall effective external field $\bm{B}_l$ of layer $l$ to be: 
    \begin{equation}
    -\gamma \bm{m}_{l} \times  \bm{B}_l\equiv -\gamma_l \bm{m}_{l} \times  \bm{B}_{ext}+\bm{\tau}_{\mathrm{F L},l} 
    \end{equation}
    where $\bm{B}_{\mathrm{ext}}=\bm{B}-\bm{B}_{\mathrm{ex}}-\bm{B}_{\mathrm{d}}-... = \mu_0\bm H_{\mathrm{ext}}$ is the effective field's external (applied) part. Here  
    $\bm B_{\mathrm{ex}}$ is the short-range exchange field,
    $\bm B_{\mathrm d}$ the long-range demagnetizing (dipolar) field,
    and the ellipsis stands for further internal terms
    (anisotropy, DMI, \dots).

    In the static case, we replace $\partial/\partial t$ with $-i\omega$ and write everything into a matrix. We directly solve the eigenvalue problem. It is equivalent to solving the non-Hermitian Schrödinger-like equation in second quantization if we cancel the Gilbert damping term by a damping-like torque  $\bm{\tau}_{\mathrm{D L},l}$. 
    Explicitly, the mappings between the LLG equation and second quantization are:  
    \begin{subequations}\label{eq:parameter-matching}
    \begin{align}
    M_l &\;=\; \frac{g\,\mu_\mathrm{B}\,S_l}{a_0^{3}},
    \label{eq:Ms}\\[6pt]
    g\,\mu_\mathrm{B} &= \hbar\,\gamma_l,
    \label{eq:gmuB}\\[6pt]
    A_l^{\mathrm{ex}} &= \frac{2J_l\,S_l^{2}}{a_0}.
    \label{eq:Aex}
    \end{align}
    \end{subequations}
    where $\gamma_l$ is electron’s gyromagnetic ratio of atom $l$, $A_{l}^{\mathrm{ex}}$ is the exchange stiffness constant, $\mu_B$ is the Bohr magneton and $g$ is the Landé factor. 
    Combining Eqs.\;(\ref{eq:Ms})–(\ref{eq:Aex}) one obtains the identity that
    relates the Heisenberg coupling \(J_l\) to the coefficient that appears in
    the exchange term of the LLG equation:
    \begin{equation}
    \frac {J_l S_l a_0^2}{\hbar} = \frac{2\gamma_l A_{l}^{\mathrm{ex}}}{M_l}
    \end{equation}

    In simulation, we have $A_{a}^{\mathrm{ex}} = 3.7\times 10^{-12}\ \mathrm{J/m}$,
    $A_{b}^{\mathrm{ex}} = 8.7\times 10^{-12}\ \mathrm{J/m}$, 
    $M_a = 1.4 \times 10^{5}\ \mathrm{A/m}$, 
    $M_b = 7.4 \times 10^{5}\ \mathrm{A/m}$, 
    $\gamma_a = \gamma_b = 2\pi \times 28 \ \mathrm{GHz/T}$, and 
    $d_a=d_b  = 10\ \mathrm{nm}$

    \newcommand{\csl}[1]{\texttt{#1}}   
    
    In Fig.\ref{fig:11}, we present the simulation process used to model dipolar interactions between magnetic layers by coupling the \csl{Micromagnetics and AC/DC Modules}. As an example, we demonstrate the procedure for an FM bilayer; simulations for other systems follow the same methodology. Importantly, our simulation approach does not rely on the nearest-neighbor approximation for dipolar interactions.

    The simulation employs two \csl{Micromagnetics Frequency Domain Modules} (MMF1 and MMF2), each representing a distinct FM insulator, as shown in Fig.\ref{fig:11}(a)-(b). The intrinsic properties of each FM insulator—such as exchange stiffness, gyromagnetic ratio, and saturation magnetization—are specified within the \csl{LLG Equation} settings. Both \csl{static external fields} (SEF) and \csl{dynamic external fields} (DEF) are included: the SEF corresponds to the effective magnetic field experienced by the layer ($B^y$) and accounts for both spin-orbit torque and any applied external field, while the DEF encompasses the dipolar interaction exported from the MF module as well as the excitation field.

    For band structure simulations, we apply a localized uniform excitation at the center of the layer, approaching a delta function to generate all Fourier components. As indicated by the pink region in Fig.\ref{fig:11}(a), this excitation is given by $\bm{h_1}$. For spin wave transmission spectroscopy simulations, a spatially uniform excitation with fixed wave vector and frequency is used, represented by the uniform blue region. This excitation is given by: $\bm{h}_2 = h_0 \exp(-ik_0x)$. The excitation formulas are shown below with $h_0 = 8000\ \mathrm{A/m}$:
    \begin{equation}
    \bm{h_1} = \begin{cases}
     h_0\hat{x}\ (x\in\ [-0.5 \ \mathrm{nm},0.5\ \mathrm{nm}] ) \\
    0\ (\mathrm{Otherwise})
    \end{cases} 
    \end{equation}
    \begin{equation}
    \bm{h_2} = h_0 e^{(-ik_0x)} \hat{z}
    \end{equation}

    The initial magnetization $\bm{m}_{a/b/c,0}$ is set along $\pm\hat{y}$ to represent parallel and antiparallel configurations. PBCs are applied along the $x$ direction for all layers.

    To simulate dipolar interactions, the \csl{AC/DC Magnetic Fields Module} (MF) is coupled to MMF1 and MMF2 by creating two separate \csl{Ampère’s Law} domains, as specified in Fig.\ref{fig:11}(a). The dynamic unit magnetization from the MMF modules is input to the MF module, which, in turn, calculates the corresponding magnetic fields. These fields are subsequently incorporated into the DEF in the MMF modules. This iterative exchange of magnetization and dynamic magnetic field enables the simulation of both interlayer and intralayer dipolar interactions. To ensure accuracy, an additional \csl{Ampère’s Law} setting is defined for the surrounding air domain. The outer boundary of the air domain utilizes the \csl{Magnetic Insulation} condition. Finally, all domains' initial magnetic potentials $\bm{A}$ in MF are set to zero at the start of the finite element simulation.

    The full simulation workflow is depicted in Fig.\ref{fig:11}(c), with processes performed in MATLAB highlighted in orange. Figs.\ref{fig:11}(a)-(b) show the geometry, mesh, and relevant physics settings. The mesh for the magnetic insulator domains is \csl{mapped type}, in order to mimic a simple cubic lattice. It is consistent with the theoretical assumptions and facilitates efficient Fourier-Transform-based post-processing in MATLAB.
    
    To hold the non-equilibrium state in simulation, we introduce a damping-like torque to both layers to overcome  Gilbert damping. For simplicity, we only simulate one single bilayer unit of the FM case. This section contains only the antiparallel case. We use a linearly polarized spin wave pulse to excite the bilayer at the left edge to generate a wide range of $k_x$ and study the intrinsic properties of the system. Then we do the Fourier Transform to magnetic moments $m_l$ along the $x$ direction to obtain the band structure in layer $l$. The results are shown in Fig. \ref{fig:6}. Specifically, when including the intralayer dipolar interaction, in Fig. \ref{fig:6}(d), the originally trivial state becomes nontrivial by the occurrence of the anticrossing only when $k_x<0$ and $\omega >0$ (and $k_x>0$ and $\omega <0$ due to PHS). The simulation and calculation results are consistent, indicating the feasibility of realizing such a non-equilibrium state and the consequent new non-trivial surface states.

    As for the parallel case, we simulate with the same parameters except changing the alignment of the magnetization in each layer to the same direction. The simulation band structure results are shown in Fig. \ref{fig:3}(b)-(c).

    To show that the surface state is detectable in experiment, we further simulate propagating spin-wave spectroscopy. We instead use a linearly polarized magnetic field to excite the bilayer at $k_x = \pm k_0$ to see the response of the system. $+k_0$ gives the non-trivial surface state between antimagnon \(a\) and \(c\). We calculate the transmission rate as: 
    \begin{equation}
    |T_l|^2 =\log _{10}\left(\frac{\int_{0}^{L_0}\left|m_{l}^y\right|^2 \mathrm{d}x}{L_0\left(m_{l,\mathrm{max}}^y\right)^2}\right)
    \end{equation} 
    where $L_0$ is the length of bilayer in $x$ direction and $m_{l,\mathrm{max}}^y$ is the maximum $m_l^y$ among different excitation. In Fig. \ref{fig:7}(a), there is a peak in the transmission rate around $5\ \mathrm{GHz}$ with $+k_0$  excitation but not with $-k_0$. Similarly, in Fig. \ref{fig:7}b, there is a peak in the transmission rate around $10\ \mathrm{GHz}$ with $-k_0'$ but not with $+k_0'$, indicating that the topological nonreciprocal surface states exist and are detectable. Specifically, the case in Fig. \ref{fig:7}(b) should be trivial when not considering intralayer interaction in the previous research. Here it is non-trivial, proving that dipolar magnon-antimagnon interaction indeed influences the topological order.

    We further analyze the chirality of this model. Recall that magnon refers to RH precession of the magnetic moment and antimagnon to LH precession. In isolated FM insulators, RH chirality is naturally stable, while LH chirality is destabilized due to Gilbert damping \cite{Coey}. Previous studies explored AFM coupling between two FM insulators, using coupled LLG equations \cite{AFM_coupling_1, AFM_coupling_2, AFM_coupling_3}. They identified eigenstates where each layer exhibits a $\pi$ phase shift in dynamic magnetic moments, with a $\pi/2$ or $-\pi/2$ phase shift between the $x$ and $z$ components, enabling both LH and RH precessions. Our findings reveal that four chirality combinations in the bilayer system —RH-LH, LH-RH, LH-LH, and RH-RH—are possible. Notably, RH-LH and LH-RH configurations cannot be achieved through AFM coupling alone \cite{Chirality1_Ferri_four}. Our approach, diverging from the perpendicular angular momentum transfer of the Kittel mode \cite{Chirality2_comments}, allows spin angular momentum with specific chirality to propagate longitudinally through the film, introducing a new degree of freedom for engineering magnetic systems.
        
    We examine four particular states in the bilayer unit, which are marked by purple circles in the subplot of Fig \ref{fig:6}(c). We use linear excitation and simulate the chiral response of these four states.   
    The real space spin directions with respect to time are depicted in Fig. \ref{fig:8}(a)-(d), where plots are in the $xz$ plane. Precessioning magnetic moments are shown with black arrows, while green and blue circular arrows represent RH and LH modes, respectively, illustrating the precessioning direction in each layer. The simulation results are consistent with the eigenvalue results from calculating band structures. Note that we cannot compare phases between different eigenvalues due to gauge variance. However, we can compare the phase difference between different components within the same eigenvector. 
    
    As shown in Fig. \ref{fig:8}(c), the chirality configuration is protected by the topological surface state, resulting in decoupled intrinsic states. By tuning external fields and torques, any of the four chirality configurations can be engineered to be protected by topological surface states.
    
    The transmission rates of RH and LH are plotted in Fig. \ref{fig:9}. By comparing the magnitude difference of RH and LH transmission rate between the same type of spin, we claim that these states indeed are nearly pure states with indicated chirality.

\end{document}